\documentclass[11pt,bibliography=totoc]{scrartcl}

\usepackage{soul}
\usepackage[utf8]{inputenc}
\usepackage[english]{babel}
\usepackage{graphicx,amsmath,amsfonts,amssymb,dcolumn,amsthm,slashed,bbm,xspace,pgffor,tikz,mathbbol,latexsym,cite,braket}
\usepackage{microtype}
\usepackage{lmodern}
\usepackage[normalem]{ulem}
\usepackage{xcolor}
\usepackage[colorlinks=true,citecolor=blue,urlcolor=blue,linkcolor=blue]{hyperref}
\usepackage[hang]{footmisc} 
\usepackage[affil-it]{authblk}
\usepackage[compat=1.1.0]{tikz-feynhand}
\usepackage{multicol}

\numberwithin{equation}{section}

\begin{document}

\title{\textbf{Correspondence between the twisted $N = 2$ super-Yang-Mills and conformal Baulieu-Singer theories }}

\author{Octavio C.~Junqueira$^{a,b}$\thanks{\href{mailto:octavioj@pos.if.ufrj.br}{octavioj@pos.if.ufrj.br}}~ and  Rodrigo F.~Sobreiro$^b$\thanks{\href{mailto:rodrigo\_sobreiro@id.uff.br}{rodrigo\_sobreiro@id.uff.br}} }
\affil{\footnotesize $^{a}$ UFRJ --- Universidade Federal do Rio de Janeiro, Instituto de Física,\\
Caixa Postal 68528, Rio de Janeiro, Brasil\\
 $^{b}$ UFF --- Universidade Federal Fluminense, Instituto de F\'isica,\\ Av.~Litoranea s/n, 24210-346, Niter\'oi, RJ, Brasil}

\date{}
\maketitle

\begin{abstract}
We characterize the correspondence between the twisted $N=2$ super-Yang-Mills theory and the Baulieu-Singer topological theory quantized in the self-dual Landau gauges. While the first is based on an on-shell supersymmetry, the second is based on an off-shell Becchi-Rouet-Stora-Tyutin symmetry. Because of the equivariant cohomology, the twisted $N=2$ in the ultraviolet regime and Baulieu-Singer theories share the same observables, the Donaldson invariants for 4-manifolds. The triviality of the Gribov copies in the Baulieu-Singer theory in these gauges shows that working in the instanton moduli space on the twisted $N=2$ side is equivalent to working in the self-dual gauges on the Baulieu-Singer one. After proving the vanishing of the $\beta$ function in the Baulieu-Singer theory, we conclude that the twisted $N=2$ in the ultraviolet regime, in any Riemannian manifold, is correspondent to the Baulieu-Singer theory in the self-dual Landau gauges---a conformal gauge theory defined in Euclidean flat space. 
\end{abstract}

\newpage
\tableofcontents
\newpage

\section{Introduction}

Throughout the 1980s, based on the self-dual Yang-Mills equations introduced by A. Belavin, A. Polyakov, A. Schwartz, and Y. Tyupkin in their study of instantons \cite{Belavin:1975fg},  S. K. Donaldson discovered and described topological structures of polynomial invariants for smooth 4-manifolds \cite{DONALDSON1990257, Donaldson:1983wm, donaldson1987}. The connection between the Floer theory for 3-manifolds \cite{floer1987, floer1988} and Donaldson invariants for 4-manifolds with a nonempty boundary, \textit{i.e.}, that assumes values in Floer groups, has led to Atiyah's conjecture \cite{Atiyah:1987ri, Witten:1988ze}. In this conjecture, he proposed that the Floer homology must lead to a relativistic quantum field theory. This conjecture was the motivation for Witten's topological quantum field theory (TQFT) in four dimensions, as Witten himself admits \cite{Witten:1988ze}. In \cite{Atiyah:1987ri}, Atiyah showed that Floer's results \cite{floer1988} can be seen as a version of a supersymmetric gauge theory. Answering Atiyah's conjecture, Witten found a relativistic formulation of \cite{Atiyah:1987ri}, capable of reproducing the Donaldson polynomials in the the weak coupling limit of the twisted $N=2$ SYM theory. This TQFT is commonly referred to as the Donaldson-Witten theory (DW) in the Wess-Zumino gauge \cite{West:1990}.

In practice, TQFTs have the power to reproduce topological invariants of the basis manifold as observables. The first one to obtain topological invariants from a quantum field theory was A. S. Schwarz in 1978 \cite{Schwarz:1978cn}. He showed that the Ray-Singer analytic torsion \cite{Ray:1973sb} can be represented
as a partition function of the Abelian Chern-Simons (CS) action, which is invariant by diffeomorphisms. The Schwarz topological theory was the prototype of Witten theories in the 1980s. Indeed, the well-known Witten paper in which he reproduces the Jones polynomials of knot theory \cite{Witten:1988hf}, is the non-Abelian generalization of Schwarz`s results \cite{Schwarz:1978cn}. In this work, Witten is actually able to represent topological invariants of 3-manifolds as the partition function of the non-Abelian CS theory.

After Witten's result \cite{Witten:1988ze}, L.~Baulieu and I.~M.~Singer (BS) showed in \cite{Baulieu:1988xs} that the same topological observables can be obtained from a gauge-fixed topological action. In such an approach, the Becchi-Rouet-Stora-Tyutin (BRST) symmetry \cite{Becchi:1975nq,Tyutin:1975qk,Piguet:1995er} plays a fundamental role. It is not built through a linear transformation of a supersymmetric gauge theory, like Witten's TQFT. It is built through a gauge-fixing procedure of a topological-invariant action, in such a way that the BRST operator naturally appears as nilpotent without requiring the use of equations of motion. The geometric interpretation of the BS theory is that the non-Abelian topological theory lies in a universal space graded as a sum of the ghost number and the form degree, where the vertical direction of this double complex is determined by the ghost number and the horizontal one is determined by the form degree. In this space, the topological BRST transformations is written in terms of a universal connection, and its curvature naturally explains the BS approach as a topological Yang-Mills theory with the same global observables of Witten's TQFT. 

From the physical point of view, the motivation to study TQFTs comes from the mathematical tools of such theories, capable of revealing the topological structure of field theories that are independent of variations of the metric, and consequently of the background choice. One of the major obstacles to constructing a quantum theory of gravity is the integration over all metrics. The introduction of a topological phase in gravity would have the power to make a theory of gravity arise from a symmetry breaking mechanism of a background independent topological theory\footnote{We must say that the introduction of such a topological phase is one of the intricate problems in topological quantum field theories, since one should develop a mechanism to break the topological symmetry.} \cite{Witten:1988ze, vanBaal:1989aw}. On the other hand, we can investigate conformal properties of field theories via topological models. In three dimensions, for instance, the connection between the three-dimensional Chern-Simons theory and two-dimensional conformal theories is well known \cite{Witten:1988hf}. In four dimensions, TQFTs are intimately connected with the AdS/CFT correspondence \cite{Witten:1998wy, BenettiGenolini:2017zmu}. More recently, motivated by string dualities, a topological gravity phase in the early Universe was proposed \cite{Agrawal:2020xek}. Such a phase could explain some puzzles concerning early Universe cosmology.  

The fact that DW theory at the UV regime and BS theories share the same observables is a well-known result in the literature \cite{Witten:1988ze,Baulieu:1988xs, Weis:1997kj, Delduc:1996yh, Boldo:2003jq}. In this paper, we characterize the correspondence between DW TQFT and a conformal BS gauge theory at quantum level. While Witten's theory is based on the \textit{twisted} version of the $N=2$ super-Yang-Mills theory, the mentioned conformal theory is based on the Baulieu-Singer BRST gauge-fixing approach to a topological action \cite{Baulieu:1988xs}. In recent works \cite{Junqueira:2017zea, Junqueira:2018zxr, Junqueira:2018xgl}, the existence of an extra bosonic symmetry was proved in the case of self-dual Landau gauges\footnote{For simplicity we will refer to the (anti-)self-dual Landau gauges, defined by instantons and anti-instantons configurations, see gauge condition \eqref{gf3}, only by the denotation \textit{self-dual gauges}.}. This bosonic symmetry relates the Faddeev-Popov and the topological ghost fields. Together with the known vector supersymmetry \cite{Brandhuber:1994uf} and the vanishing three-level gauge propagator, one observes that the BS theory at the self-dual Landau gauges is indeed tree-level exact \cite{Junqueira:2018xgl}. Essentially, the proof of this property is diagrammatic with some help of algebraic renormalization techniques \cite{Piguet:1995er}. This remarkable property inevitably implies a vanishing $\beta$ function, since it does not receive quantum corrections. Nevertheless, an entire algebraic proof was still lacking until now. It turns out that, for a complete proof of the vanishing of the $\beta$ function of the BS theory in the self-dual Landau gauges, one extra property must be considered: the fact that the Gribov copies are inoffensive to the self-dual BS theory \cite{Dudal:2019bjh}. This property establishes that the self-dual BS theory is conformal, as it allows to recover some discrete symmetries. The use of these symmetries makes it possible to eliminate the renormalization ambiguities discussed in \cite{Junqueira:2018zxr}. With this information, we were able to establish the correspondence between self-dual BS theories (a conformal gauge theory defined in Euclidean spaces) for any value of the coupling constant and DW theory at the deep UV. 

The paper is organized as follows. Section \ref{TQFT} contains an overview of the main properties of DW and BS theories. We introduce the main aspects of each approach, explaining how each one is constructed from different quantization schemes. As the quantum properties of the Witten's TQFT is well known in literature, we dedicate Section \ref{SDLgauges} to discussing the quantum properties of the BS theory in the self-dual Landau gauges. In Section \ref{Beta}, we analyze and compare the corresponding $\beta$ functions of each model, after proving that the self-dual BS is conformal. Finally, in Section \ref{DW/BS}, we describe the quantum correspondence between Witten and self-dual BS topological theories. Section \ref{Conclusions} contains our concluding remarks.

\section{Topological quantum field theories}\label{TQFT}

A topological quantum field theory on a smooth manifold is a quantum field theory  which is independent of the metric on the basis manifold. Such a theory has no dynamics, no local degrees of freedom, and is only sensitive to topological invariants which describe the manifold in which the theory is defined. The observables of a TQFT are naturally metric independent. The latter statement leads to the main property of topological field theories, namely, the metric independence of the observable correlation functions of the theory, 
\begin{equation} \label{gindependence}
\frac{\delta}{\delta g_{\mu\nu}} \langle  \mathcal{O}_{\alpha_1}(\phi_i) \mathcal{O}_{\alpha_2}(\phi_i)  \cdots \mathcal{O}_{\alpha_p}(\phi_i)\rangle = 0\;,
\end{equation}
with
\begin{equation}
\langle  \mathcal{O}_{\alpha_1}(\phi_i) \mathcal{O}_{\alpha_2}(\phi_i)  \cdots \mathcal{O}_{\alpha_p}(\phi_i)\rangle = \mathcal{N}\int \mathcal[{D}\phi_i]   \mathcal{O}_{\alpha_1}(\phi_i) \mathcal{O}_{\alpha_2}(\phi_i)  \cdots \mathcal{O}_{\alpha_p}(\phi_i) \,e^{- S[\phi]}\;,
\end{equation}
where $g_{\mu\nu}$ is the metric tensor, $\phi_i(x)$ are quantum fields, $\mathcal{O}_\alpha$ the functional operators of the fields composing global observables, $S[\phi]$ is the classical action, and $\mathcal{N}$ the appropriate normalization factor. A typical operator $\mathcal{O}_\alpha$ is integrated over the whole space in order to capture the global structures of the manifold. Since there are no particles, the only nontrivial observables are of global nature \cite{Sorella:1989ri, Blasi:1989ka}.  

As a particular result of \eqref{gindependence}, the partition function of a topological theory is itself a topological invariant, 
\begin{equation} \label{deltagZ}
    \frac{\delta}{\delta g_{\mu\nu}} Z[J] = 0\;,
\end{equation}
insofar as $Z[J]$ represents the expectation value of the vacuum in the presence of a external source, $Z[J] = \langle 0 \vert 0\rangle_J$. As discussed in \cite{Labastida:1997pb}, if the action is explicitly independent of the metric, the topological theory is said to be of \textit{Schwarz type}; otherwise, if the variation of the action with respect to the metric gives a ``BRST-like"-exact term, one says the theory is of \textit{Witten type}. More precisely, being $\delta$ an infinitesimal transformation that denotes the symmetry of the action $S$ which characterizes the observables of the model, then, if the following properties are satisfied, 
\begin{equation} \label{topoconditions}
\delta \mathcal{O}_\alpha(\phi_i) = 0\,, \quad T_{\mu\nu}(\phi_i) = \delta G_{\mu\nu}(\phi_i)\;,
\end{equation}
where $T_{\mu\nu}$ is the energy-momentum tensor of the model,  
\begin{equation}\label{STmunu}
\frac{\delta}{\delta g_{\mu\nu}} S = T_{\mu\nu}\;,    
\end{equation}
and $G_{\mu\nu}$ some tensor, then the quantum field theory can be regarded as topological. Obviously, in this case eq. \eqref{deltagZ} is also satisfied, since the expectation value of the $\delta$-exact term vanishes\footnote{The nilpotent $\delta$-operator works precisely as a BRST operator, and it is well-known that expectation values of BRST-exact terms vanish. For a further analysis concerning renormalization properties, and the definition of physical observables, see \cite{Piguet:1995er,Becchi:1975nq,Kugo:1979gm}.} \cite{Witten:1988ze, Baulieu:1988xs}. In fact, by using \eqref{topoconditions} and \eqref{STmunu}, and assuming that the measure $[{D}\phi_i]$ is invariant under $\delta$,
\begin{eqnarray} \label{toporesult}
\frac{\delta}{\delta g_{\mu\nu}} \langle  \mathcal{O}_{\alpha_1}(\phi_i) \mathcal{O}_{\alpha_2}(\phi_i)  \cdots \mathcal{O}_{\alpha_p}(\phi_i)\rangle &=& - \int \mathcal[{D}\phi_i]   \mathcal{O}_{\alpha_1}(\phi_i) \mathcal{O}_{\alpha_2}(\phi_i)  \cdots \mathcal{O}_{\alpha_p}(\phi_i) T_{\mu\nu} \,e^{- S} \nonumber \\
&=&  \langle \delta[ \mathcal{O}_{\alpha_1}(\phi_i) \mathcal{O}_{\alpha_2}(\phi_i)  \cdots \mathcal{O}_{\alpha_2}(\phi_i)G_{\mu\nu}]\rangle \nonumber \\
&=& 0\;. 
\end{eqnarray}
In the above equation we assumed that all $\mathcal{O}_\alpha$ are metric independent. Nevertheless this is not a requirement of the theory. It is also possible to have a more general theory in which
\begin{equation}\label{observablesdeltaexact}
\frac{\delta}{\delta g_{\mu\nu}} \mathcal{O}_\alpha = \delta \mathcal{Q}_{\mu\nu} \neq 0\;,
\end{equation}
that preserves the topological structure of $\delta_{g_{\mu\nu}} \langle \mathcal{O}_{\alpha_1} \cdots \mathcal{O}_{\alpha_p} \rangle = \langle \delta(\cdots) \rangle = 0$ \cite{Labastida:1997pb}. Analogously to the BRST operator, eq. \eqref{toporesult} only makes sense if the $\delta$ operator is nilpotent\footnote{In Donaldson-Witten theory, for instance, such an operator is on-shell nilpotent, \textit{i.e.}, $\delta^2 = 0$ by using the equations of motion.}.

\subsection{Donaldson-Witten theory}\label{DWtheory}

As mentioned in the introduction, Witten constructed in \cite{Witten:1988ze} a four-dimensional generalization of  \cite{Atiyah:1987ri}, capable of reproducing the Donaldson invariants \cite{DONALDSON1990257, Donaldson:1983wm, donaldson1987} in the weak coupling limit. Such a construction can be obtained from the twist transformation of the $N=2$ SYM. Let us quickly revise some important features of such approach.

\subsubsection{The twist transformation}
\label{twist}


The eight supersymmetric charges ($Q^i_{\alpha}, \,\bar{Q}_{j\dot{\alpha}}$) of $N=2$ SYM theories obey the SUSY algebra
\begin{eqnarray}
\{Q^i_{\alpha}, \,\bar{Q}_{j\dot{\alpha}}\} &=& \delta^i_j (\sigma_\mu)_{\alpha \dot{\alpha}} \partial_\mu \;,\nonumber \\
\{Q^i_{\alpha},\,{Q}_{j{\alpha}}\} &=& \{\bar{Q}^i_{\dot{\alpha}}, \,\bar{Q}_{j\dot{\alpha}}\} = 0\;,
\end{eqnarray}
where all indices $(i,j, \alpha, \dot{\alpha})$ run from one to two. The indices $(i,j)$ denote the internal $SU(2)$ symmetry of the $N=2$ SYM action, and $(\alpha, \dot{\alpha})$ are Weyl spinor indices: $\alpha$ denotes right-handed spinors, and $\dot{\alpha}$, left-handed ones. The fact that both indices equally run from one to two suggests the identification between spinor and supersymmetry indices, 
\begin{equation} \label{identification}
    i \equiv \alpha \;. 
\end{equation}
The $N=2$ SYM action theory possesses a gauge group symmetry given by
\begin{equation}
    SU_L(2) \times SU_R(2) \times SU_I(2) \times U_{R}(1)\;, 
\end{equation}
where $SU_L(2) \times SU_R(2)$ is the rotation group, $SU_I(2)$ is the internal supersymmetry group labeled by $i$, and $U_{{R}}(1)$, the so-called $\mathcal{R}$-symmetry defined by the supercharges ($Q^i_{\alpha}, \,\bar{Q}_{j\dot{\alpha}}$) which are assigned eigenvalues $(+1$, $-1)$, respectively. The identification performed in eq. \eqref{identification} amounts to a modification of the rotation group,
\begin{equation}
SU_L(2) \times SU_R(2) \rightarrow SU_L(2) \times SU_R(2)^\prime\;,
\end{equation}
where $SU_R(2)^\prime$ is the diagonal sum of $SU_R(2)$ and $SU_I(2)$. The \textit{twisted} global symmetry of $N=2$ SYM takes the form $SU_L(2) \times SU_R(2)^\prime \times U_R(1)$, with the corresponding \textit{twisted} supercharges
\begin{equation}
Q^i_\alpha \rightarrow Q^\beta_\alpha\,, \quad \bar{Q}_{i\bar{\alpha}} \rightarrow \bar{Q}_{\alpha\dot{\alpha}}\;,
\end{equation}
which can be rearranged as
\begin{eqnarray} 
\frac{1}{\sqrt{2}}\epsilon^{\alpha\beta} Q_{\alpha\beta} &\equiv& \delta\;,\label{Qdelta} \\
\frac{1}{\sqrt{2}} \bar{Q}^{\alpha\dot{\alpha}}(\sigma_\mu)^{\dot{\alpha}\alpha}&\equiv& \delta_\mu \;, \label{Qdeltamu}\\
\frac{1}{\sqrt{2}} (\sigma_{\mu\nu})^{\dot{\alpha}\alpha}Q_{\dot{\alpha}\alpha}&\equiv& d_{\mu\nu} \;, \label{Qdeltamunu}
\end{eqnarray}
where we adopt the conventions for $\epsilon^{\alpha\beta}$, $(\sigma^\mu)^{\alpha\dot{\alpha}}$ and $(\sigma_{\mu\nu})^{\dot{\alpha}\alpha}$ as the same of \cite{Wess:1992cp}. The operator $d_{\mu\nu}$ is manifestly self-dual due to the structure of $\sigma_{\mu\nu}$, 
\begin{equation}
d_{\mu\nu} = \frac{1}{2} \varepsilon_{\mu\nu\lambda\rho} d^{\lambda\rho}\;,
\end{equation}
reducing to three the number of its independent components. The operators $\delta$, $\delta_\mu$ and $\delta_{\mu\nu}$ possess eight independent components in which the eight original supercharges $(Q_{\beta\alpha},\,\bar{Q}_{\alpha\dot{\alpha}})$ are mapped into. These operators obey the following \textit{twisted} supersymmetry algebra
\begin{eqnarray}
\delta^2 &=& 0\;,\label{delta2}\\
\{\delta, \delta_\mu\} &=& \partial_\mu\;,\label{deltapartial}\\
\{\delta_\mu, \delta_\nu\} &=& \{d_{\mu\nu}, \delta\} =  \{ d_{\mu\nu}, d_{\lambda\rho}\}=0\;,\\
\{\delta_{\mu}, d_{\lambda\rho}\} &=&  - ( \varepsilon_{\mu\lambda\rho\sigma}\partial^\sigma + g_{\mu\lambda}\partial_\rho - g_{\mu\rho}\partial_\lambda )\;.
\end{eqnarray}
The nilpotent scalar supersymmetry charge $\delta$ defines the cohomology of Witten's TQFT, as its observables appear as cohomology classes of $\delta$, which is invariant under a generic differential manifold. It is implicit in the anti-commutation relation \eqref{deltapartial} the topological nature of the model, as it allows to write the common derivative as a $\delta$-exact term. 

The gauge multiplet of the $N=2$ SYM in Wess-Zumino gauge is given by the fields
\begin{equation}
(A_\mu, \psi^i_\alpha, \bar{\psi}^i_{\dot{\alpha}}, \phi, \bar{\phi})\;,
\end{equation}
where $\psi^i_\alpha$ is a Majorana spinor (the supersymmetric partner of the gauge connection $A_\mu$), and $\phi$, a scalar field, all of them belonging to the adjoint representation of the gauge group. The \textit{twist transformation} is defined by the identification eq. \eqref{identification}, and thus only acts on the fields $(\psi^i_\mu, \bar{\psi}^i_\mu)$, leaving the bosonic fields $(A_\mu, \phi, \bar{\phi})$ unaltered. Explicitly, the \textit{twist transformation} is given by the linear transformations\footnote{Notation: $\Phi_{(\alpha\beta)} = \Phi_{\alpha\beta} + \Phi_{\beta\alpha}$ and $\Phi_{[\alpha\beta]} =\Phi_{\alpha\beta} - \Phi_{\beta\alpha}$.}
\begin{eqnarray}
\psi^i_\beta \,\,&\rightarrow&\,\, \psi_{\alpha\beta} = \frac{1}{2}\left(\psi_{(\alpha\beta)} + \psi_{[\alpha\beta]}\right) \;, \label{twist1} \\
\bar{\psi}^i_{\dot{\alpha}} &\rightarrow& \bar{\psi}_{\alpha\dot{\alpha}} \,\,\rightarrow\,\, \psi_\mu = (\sigma_\mu)^{\alpha\dot{\alpha}} \bar{\psi}_{\alpha\dot{\alpha}}\,,
\end{eqnarray}
together with
\begin{eqnarray}
\psi_{(\alpha\beta)} \,\, &\rightarrow& \,\, \chi_{\mu\nu}  = (\sigma_{\mu\nu})^{\alpha\beta}\psi_{(\alpha\beta)}\;, \\
\psi_{[\alpha\beta]} \,\, &\rightarrow & \eta  = \varepsilon^{\alpha\beta} \psi_{[\alpha\beta]}\;.
\end{eqnarray}
The \textit{twist} consists of a mapping of degrees of freedom. The field $\bar{\psi}_{\alpha\dot{\alpha}}$ has four independent components as $(\alpha, \dot{\alpha}) = \{1,2\}$, and is mapped into the field $\psi_\mu$ that also has four independent components of the path integral, as the Lorentz index $\mu = \{1,2,3,4\}$ in four dimensions. In the other mappings the same occurs, as the symmetric part of $\psi_{\alpha\beta}$, \textit{i.e.}, $\psi_{(\alpha\beta)}$ has three independent components mapped into the self-dual field $\chi_{\mu\nu}$, and the antisymmetric part, $\psi_{[\alpha\beta]}$, with only one independent component, into $\eta$, a scalar field. We must note that $(\psi_\mu, \chi_{\mu\nu}, \eta)$ are anti-commuting field variables due to their spinor origin. 

Because it is a linear transformation, the \textit{twist} simply corresponds to a change of variables with trivial Jacobian that could be absorbed in the normalization factor, in other words, both theories (before and after the twist) are perturbatively indistinguishable. Finally, twisting the $N=2$ SYM action ($S^{N=2}_{SYM}$) \cite{Witten:1988ze, Blasi:2000qw}, in flat Euclidean space, we obtain the Witten four-dimensional topological Yang-Mills action ($S_{W}$), 
\begin{equation}
S^{N=2}_{SYM}[A_\mu, \psi^i_\alpha, \bar{\psi}^i_{\dot{\alpha}}, \phi, \bar{\phi}] \,\, \rightarrow \,\, S_{W}[A_\mu, \psi_\mu, \chi_{\mu\nu}, \bar{\phi}, \phi]\;, 
\end{equation} 
where
\begin{eqnarray} \label{SWitten}
S_W &=& \frac{1}{g^2} \text{Tr} \int d^4x \, \left( \frac{1}{2} F^{+}_{\mu\nu} {F^+}^{\mu\nu} - \chi_{\mu\nu}\left(D_\mu\psi_\nu - D_\nu \psi_\mu\right)^+ + \eta D_\mu \psi^\mu \right.\nonumber\\
&-&\left.\frac{1}{2} \bar{\phi} D_\mu D^\mu \phi + \frac{1}{2}\bar{\phi}\{\psi_\mu, \psi_\mu\}-\frac{1}{2}\phi \{\chi_{\mu\nu}, \chi_{\mu\nu}\} -\frac{1}{8} \left[\phi, \eta\right] \eta \right. \nonumber\\
&-&\left.\frac{1}{32} \left[\phi, \bar{\phi}\right] \left[\phi, \bar{\phi}\right]\right)\;,
\end{eqnarray}
wherein $F^+_{\mu\nu}$ is the self-dual field\footnote{Following \cite{Witten:1988ze, Blasi:2000qw}, we are considering the positive sign, that corresponds to anti-instantons in the vaccum. A similar construction can be done for instantons, only by changing the sign.}
\begin{equation} 
\label{F^+}
F^+_{\mu\nu} = F_{\mu\nu}+ \widetilde{F}_{\mu\nu}\,, \quad (\widetilde{F}^+_{\mu\nu} = F^+_{\mu\nu})\;,
\end{equation}
with $\widetilde{F}_{\mu\nu} = \frac{1}{2} \epsilon_{\mu\nu\alpha\beta} F_{\alpha \beta}$, and, analogously, 
\begin{equation}
 \left(D_\mu\psi_\nu - D_\nu \psi_\mu\right)^+ = D_\mu\psi_\nu - D_\nu \psi_\mu + \frac{1}{2}\varepsilon_{\mu\nu\alpha\beta}\left(D_\alpha\psi_\beta - D_\beta \psi_\alpha\right)\;,
\end{equation}
being $D_\mu \equiv \partial_\mu - g[A_\mu, \,\cdot\,]$ the covariant derivative in the adjoint representation of the Lie group $G$, with $g$ the coupling constant. The Witten action\footnote{Technically, the Witten action \eqref{SWitten} is the four-dimensional generalization of the non-relativistic topological quantum field theory \cite{Atiyah:1987ri}, whose construction is based on the Floer theory for three-manifolds $\mathcal{M}_{3D}$, in which the Chern-Simons action is taken as a Morse function on $\mathcal{M}_{3D}$, see Floer's original paper \cite{floer1987}. In few words, the critical points of CS action ($W_{CS}$) yield the curvature free configurations, as $\frac{\delta W_{CS}}{\delta A^a_i} = - \frac{1}{2}\varepsilon^{ijk} F^{jk}$, where $F^{jk}$ is the 2-form curvature in three dimensions, which defines the gradient flows of a Morse function, see \cite{vanBaal:1989aw}. In the supersymmetric formulation of \cite{Atiyah:1987ri}, the Hamiltonian (H) is obtained via the ``supersymmetric charges" $d_t$ and $d^*_t$, from the well-known relation $d_t d_t^* + d_t^* d_t = 2H$, see \cite{Witten:1982im}, whereby $d_t = e^{-tW_{CS}}d e^{tW_{CS}}$ and $d_t^* = e^{tW_{CS}}d^*e^{-tW_{CS}}$, for a real number $t$, being $d$ the exterior derivative on the space of all connections $\mathcal{A}$, according to the transformation $\delta A^a_i = \psi^a_i$, and $d^*$ its dual. Before identifying the \textit{twist} transformation, this formulation (in four-dimensions) was employed by Witten in his original paper \cite{Witten:1988hf} to obtain the relativistic topological action \eqref{SWitten}.} \eqref{SWitten} possesses the usual Yang-Mills gauge invariance, denoted by\footnote{It is implicit in this notation the typical Yang-Mills transformations of all fields, where the gauge field transforms as $A_\mu^\prime = S^{-1} A_\mu S + S^{-1} \partial_\mu S $ with $S \in SU(N)$.}
\begin{equation} \label{gaugesymmetryWitten}
\delta^{\text{YM}}_{\text{gauge}} S_W = 0\;.
\end{equation}
The theory, however, does not possess gauge anomalies \cite{Maggiore:1994dw}. The symmetry that defines the cohomology of the theory, also known as \textit{equivariant cohomology}, is the fermionic scalar supersymmetry
\begin{eqnarray} \label{Wittensym}
\delta A_\mu &=& -\varepsilon \psi_\mu\,, \quad \delta \phi = 0\,, \quad \delta \lambda = 2i\varepsilon \eta\,, \quad \delta \eta = \frac{1}{2}\varepsilon[\phi, \bar{\phi}]\;, \nonumber\\
\delta \psi_\mu &=& - \varepsilon D_\mu \phi\,, \quad \delta \chi_{\mu\nu} = \varepsilon F^+\,,
\end{eqnarray}
where $\varepsilon$ is the supersymmetry fermionic parameter that carries no spin, ensuring that the propagating modes of commuting and anticommuting fields have the
same helicities\footnote{Precisely, the propagating modes of $A_\mu$ have helicities $(1, -1)$, and of $(\phi, \bar{\phi})$, $(0,0)$; while of the fermionic fields $(\eta, \psi, \chi)$, helicities $(1, -1, 0, 0)$.}. This symmetry relates bosonic and fermionic degrees of freedom, which are identical---an inheritance of the supersymmetric original action\footnote{The action $S_W$ is also invariant under global scaling with dimensions $(1, 0, 2,2,1,2)$ for $(A, \phi, \bar{\phi}, \eta, \psi, \chi)$, respectively; and preserves an additive $U$ symmetry for the assignments $(0, 2, -2, -1,1, -1)$. In the BRST formalism, the latter is naturally recognized as ghost numbers, as we will see later on.}. The price of working in Wess-Zumino gauge is the fact that, disregarding gauge transformations, one needs to use the equations of motion to recover the nilpotency of $\delta$ \cite{Blasi:1989ka}. This characterizes the DW theory as an \textit{on-shell approach}. One can easily verify that (see \cite{Witten:1988ze})
\begin{equation}
\delta^2 \Phi = 0\;, \quad \text{for} \quad  \Phi = \{A, \psi, \phi, \bar{\phi}, \eta\}\;,
\end{equation}
but
\begin{equation} \label{eqofmotion}
    \delta^2\chi = \text{equations of motion}\;.
\end{equation}
Considering the result of eq. \eqref{eqofmotion}, hereafter we will say that the Witten fermionic symmetry is \textit{on-shell} nilpotent. This symmetry is associated to an on-shell nilpotent ``BRST charge", $\mathcal{Q}$, according to the definition of the $\delta$ variation of any functional $\mathcal{O}$ as a transformation on the space of all functionals of field variables, namely, 
\begin{equation} \label{deltaQ}
\delta \mathcal{O} = -i\varepsilon \cdot \{\mathcal{Q}, \mathcal{O}\}\,, \quad \text{such that} \quad \mathcal{Q}^2\vert_{\textit{on-shell}} = 0\;.
\end{equation}

In order to verify that Witten theory is valid in curved spacetimes, it is worth noting that the commutators of covariant derivatives always appears acting in the scalar field $\phi$, like in $\delta Tr\{D_\mu \psi_\nu \cdot \bar{\chi}_{\mu\nu}\} = \frac{1}{2} Tr([D_\mu, D_\nu]\phi \cdot \bar{\chi}^{\mu\nu})$, so the Riemann tensor does not appear, and the theory could be extended to any Riemannian manifold. In practice one can simply take
\begin{equation}\label{Wproperty}
\int d^4x \rightarrow \int d^4x \sqrt{g}\;,
\end{equation}
in order to work in a curved spacetime. Such a theory has the property of being invariant under infinitesimal changes in the metric. This property characterizes the Witten model as a topological quantum field theory. Such a feature is verified by the fact that the energy-momentum tensor can be written as the anti-commutator
\begin{equation}
T_{\mu\nu} = \{\mathcal{Q}, V_{\mu\nu}\}\;, 
\end{equation}
which means that $T_{\mu\nu}$ is an \textit{on-shell} BRST-exact term,
\begin{equation} \label{deltaTexact}
T_{\mu\nu} =  \delta V_{\mu\nu}\,, \quad \delta^2\vert_{\textit{on-shell}}=0\;,
\end{equation}
with
\begin{eqnarray}
V_{\mu\nu} &=&  \frac{1}{2} \text{Tr}\{ F_{\mu\sigma} \chi_\nu^{\,\,\,\sigma} + F_{\nu\sigma}\chi_\mu^{\,\,\,\sigma} - \frac{1}{2} g_{\mu\nu}F_{\sigma\rho}\chi^{\sigma\rho}\} + \frac{1}{4}g_{\mu\nu} \text{Tr} \eta [\phi, \bar{\phi}] \nonumber\\
 &+&\frac{1}{2} \text{Tr} \{ \psi_\mu D^\nu \bar{\phi} + \psi_\nu D^\mu \bar{\phi} - g_{\mu\nu}\psi_{\sigma} D^{\sigma}\bar{\phi}  \} \;.
\end{eqnarray}

Equation \eqref{deltaTexact} together with $\delta S_W = 0$ means that Witten theory satisfies (on-shell) the second condition displayed in eq. \eqref{topoconditions}, that allows to say that $S_W$ automatically leads to a four-dimensional topological field model. In other words, \begin{eqnarray}\label{gmunuinv}
\frac{\delta}{\delta g_{\mu\nu}} Z_W &=& \int \mathcal{D}\Phi (- \frac{\delta}{\delta g_{\mu\nu}} \mathcal{S}_W)\text{exp}(-\mathcal{S}_W) \nonumber\\
&=& - \frac{1}{g^2} \langle \{ \mathcal{Q}, \int_M d^4x \sqrt{g}  V_{\mu\nu}\}\rangle = 0\;,
\end{eqnarray}
as all expected value of a $\delta$-exact term vanish. It remains to know which kind of topological/differential invariants can be represented by the Feynman path integral of Witten's TQFT. As we know, it will naturally reproduce the Donaldson invariants for four-manifolds. 

\subsubsection{Donaldson polynomials in the weak coupling limit}

An important feature of the twisted $N=2$ SYM is the fact that the theory can be interpreted as quantum fluctuations around classical instanton configurations. To find the nontrivial classical minima one must note that the pure gauge field terms in $S_W$ are
\begin{equation}
S^{gauge}_{W}[A] = \frac{1}{2} \text{Tr}\int d^4x \left(F_{\mu\nu}+\widetilde{F}_{\mu\nu}\right)\left(F^{\mu\nu}+\widetilde{F}^{\mu\nu}\right)\;,
\end{equation}
which is positive semidefinite, and only vanishes if the field strength $F_{\mu\nu}$ is anti-self-dual, 
\begin{equation}
F_{\mu\nu} = -\widetilde{F}_{\mu\nu} \;,
\end{equation}
the same nontrivial vacuum configuration that minimizes the Yang-Mills action in the case of anti-instantons fields. Hence, Witten's action has a nontrivial classical minima for $F = -\widetilde{F}$ and $\Phi_{\text{other fields}} = 0$. Being precise, the evaluation of the twisted $N=2$ SYM path integral computes quantum corrections to classical anti-instantons solutions. 

Another important property of Witten theory is the invariance under infinitesimal changes in the coupling constant. The variation of $Z_W$ with respect to $g^2$ yields, for similar reasons to \eqref{gmunuinv},
\begin{equation}
  \delta_{g^2} Z_W = \delta_{g^2} \left(- \frac{1}{g^2}\right)  \langle \left\{ \mathcal{Q}, X\right\} \rangle = 0\;, 
\end{equation}
where
\begin{equation}
X = \frac{1}{4}\text{Tr} F_{\mu\nu}\chi^{\mu\nu} + \frac{1}{2} \text{Tr} \psi_\mu D^\mu \bar{\phi} - \frac{1}{4} \text{Tr} \eta [\phi, \bar{\phi}]\;.
\end{equation}
The Witten partition function, $Z_W$, is independent of the gauge coupling $g^2$, therefore we can evaluate $Z_W$ in the weak coupling limit, \textit{i.e.}, in the regime of very small $g^2$, in which $Z_W$ is dominated by the classical minima. 

The instanton moduli space $\mathcal{M}_{k, N}$ is defined to be the space of all solutions to $F = \widetilde{F}$ for an instanton with a giving winding number $k$ and gauge group $SU(N)$. By perturbing $F = \widetilde{F}$ nearby the solution $A_\mu$ via a gauge transformation $A_\mu \rightarrow A_\mu + \delta A_\mu$, we obtain the self-duality equation
\begin{equation} \label{zeromodes}
D_\mu \delta A_\nu + D_\mu \delta A_\nu + \varepsilon_{\mu\nu\alpha\beta}D^\alpha \delta A^\beta = 0\;.
\end{equation}
The solutions of the above equation are called zero modes. Requiring the orthogonal gauge fixing condition\footnote{This condition is equivalent to the Landau gauge, as $D_\mu A_\mu = \partial_\mu A_\mu$. It is important to note that one can promote $\partial_\mu$ to $D_\mu$ in this case, in order to show that $A_\mu$ and $\psi_\mu$ obey the same equations.}, $D_\mu A_\mu = 0$, one gets
\begin{equation} \label{ortho}
D_\mu(\delta A_\mu) = 0\;.  
\end{equation}
The Atiyah-Singer index theorem \cite{Atiyah:1963zz, Atiyah:1978wi} counts the number of  solutions to eq. \eqref{zeromodes} and eq. \eqref{ortho}. In Euclidean spacetimes, for instance, the index theorem gives, see \cite{Tong:2005un}
\begin{equation} \label{dimmoduli}
\text{dim}(\mathcal{M}) = 4kN\;,
\end{equation}
where the modes due to global gauge transformations of the group were included. Looking at fermion zero modes, the $\chi$ equation for $S_W$ gives
\begin{equation}
D_{\mu}\psi_\nu + D_\nu \psi_\mu + \varepsilon_{\mu\nu\alpha\beta} D^\alpha\psi^\beta = 0\;,
\end{equation}
and from the $\eta$ equation, 
\begin{equation}
    D_\mu \psi^\mu = 0\;.
\end{equation}
These are the same  equations obtained for the gauge perturbation around an instanton in the orthogonal gauge fixing, so the number of $\psi$ zero modes is also given by $\mathcal{M}_{k,N}$\footnote{As Witten himself admits in his paper \cite{Witten:1988ze}, ``this relation between the fermion equations and the
instanton moduli problem was the motivation for introducing precisely this collection of fermions".}. In order to get a non-vanishing partition function, Witten assumed that the moduli space consists of discrete, isolated instantons. Precisely, he assumed that the dimension of the moduli space vanishes\footnote{Otherwise, it occurs a net violation of the $U(1)$ global symmetry of $S_W$, and $Z_W$ vanishes due to the fermion zero modes, see \cite{Witten:1988ze, tHooft:1976snw}. The dimension of the instanton moduli spaces depends on the bundle, $E$, and the manifold, $M$. In the $SU(2)$ gauge theory, it can be written as
\begin{equation}
\text{dim}(\mathcal{M}) = 8k(E) -\frac{3}{2}(\chi(M) + \sigma(M))\;,
\end{equation}
where $k(E)$ is the first Pontryagin (or winding) number of the bundle $E$, and $\chi(M)$ and $\sigma(M)$ are the Euler characteristic and signature of $M$ \cite{Atiyah:1978wi}. (For $M=R^4$, $\chi(R^4) = \sigma(R^4) = 0$.) Thus one can choose a suitable $E$ and $M$ in order to get a vanishing dimension, $\text{dim}(\mathcal{M}) = 0$.}.

In expanding around an isolated instanton, in the weak coupling limit $g^2 \rightarrow 0$, the action is reduced to quadratic terms, 
\begin{equation}
S^{(2)}_W = \int_M  d^4x \sqrt{g} \left(\Phi^{(b)} D_B \Phi^{(b)} + i\Psi^{(f)} D_F \Psi^{(f)}\right) \;,
\end{equation}
where $\Phi^{(b)} \equiv \{A, \phi, \bar{\phi}\}$ are the bosonic fields, and $\Psi^{(f)} \equiv \{\eta, \psi, \chi\}$, the fermionic ones. The Gaussian integral over all fields gives
\begin{equation}
Z_W\vert_{g^2 \rightarrow 0} = \frac{\text{Pfaff}(D_F)}{\sqrt{\det(D_B)}}\;,
\end{equation}
where $\text{Pfaff}(D_F)$ is the Pfaffian of $D_F$, \textit{i.e.}, the square root of the determinant of $D_F$  up to a sign. The supersymmetry relates the eigenvalues of the operators $D_B$ and $D_F$. The relation is a standard result in instanton calculus \cite{DAdda:1977sqj}, which yields
\begin{equation}
Z_W\vert_{g^2 \rightarrow 0} = \pm \prod_i \frac{\lambda_i}{\sqrt{\vert \lambda_i \vert}^2}\;,
\end{equation}
with $i$ running over all non-zero eigenvalues of $D_B$ $(D_F)$. Therefore, for the $k^{th}$ isolated instanton, $Z^{(k)}_W = (-1)^{n_k}$, where $n_k = 0$ or $1$ according to the orientation convention of the moduli space (Donaldson proved the orientability of the moduli space, \textit{i.e.}, that the definition of the sign of $\text{Pfaff}(D_F)$ is consistent, without global anomalies \cite{donaldson1987, Witten:1988ze}). In the end, summing over all isolated instantons, 
\begin{equation}
Z_W\vert_{g^2 \rightarrow 0} = \sum_k (-1)^{n_k}\;,
\end{equation}
which is precisely one of topological invariant for four-manifolds described by Donaldson.

The other metric independent observables are constructed in the context of eq. \eqref{observablesdeltaexact}. These observables can be generated by exploring the descent equations defined by the equivariant cohomology, \textit{i.e.}, the supersymmetry $\delta$-cohomology. For that, being $U_i$ the global charge of the operator $\mathcal{O}_i$ (see footnote 10 on page 8), it must be understood that, for the observable $\prod_i \mathcal{O}_i$, $\text{dim}(\mathcal{M}) = \sum_i U_i$\footnote{In order to construct topological invariants, the net $U$ charge must equal the dimension of the moduli space, see \cite{Witten:1988ze, vanBaal:1989aw}.}. The simplest $\delta$-invariant operator, that does not depend explicitly on the metric, and cannot be written as $\delta(X) = \{\mathcal{Q}, X\}$ (due to the scaling dimensions) is
\begin{equation}
    W_0(x) = \frac{1}{2}\text{Tr} \phi^2(x)\, , \quad U(W_0) = 4\;.
\end{equation}
Although $W_0$ is not a $\delta$-exact operator, taking the derivative of $W_0$ with respect of the coordinates, we find
\begin{equation} \label{partialW0}
\frac{\partial}{\partial x_\mu} W_0 = i \{\mathcal{Q}, \text{Tr} \phi \psi_\mu\}\;, 
\end{equation}
which is $\delta$-exact. Using the exterior derivative, $d$, we can rewrite \eqref{partialW0} as
\begin{equation}
    dW_0 = i\{\mathcal{Q}, W_1\}\;,  
\end{equation}
where $W_1$ is the 1-form $\text{Tr}(\phi\psi_\mu)dx^\mu$. A straightforward calculation gives
\begin{eqnarray}
dW_1 &=& i\{\mathcal{Q}, W_2\}\;, \quad dW_2 = i\{\mathcal{Q}, W_3\}\;, \label{deltadescent1} \\
dW_3 &=& i\{\mathcal{Q}, W_4\}\;,\quad  dW_4 = 0\;,
\end{eqnarray}
with
\begin{eqnarray}
W_2 &=& \text{Tr} (\frac{1}{2}\psi \wedge \psi + i \phi \wedge F)\;,\\ 
W_3 &=& i\text{Tr} \psi \wedge F\;, \\ 
W_4 &=& -\frac{1}{2} \text{Tr} F \wedge F\;, \label{deltadescentf}
\end{eqnarray}
where ``$\wedge$" is the wedge product, the total charge is $U = 4-k$ for each $W_k$, and $\phi, \psi$, and $F$ are zero, one, and two forms on $M$, respectively. $F$ is the field strength in the $p$-form formalism\footnote{For the definitions and conventions concerning the p-form formalism used here, see Section \ref{geometric}.}, defined in eq. \eqref{Fpform}. Considering now the integral 
\begin{equation}
I(\gamma) = \int_\gamma W_k\;,
\end{equation}
being $\gamma$ a k-dimensional homology cycle on M, we have
\begin{equation} \label{Ibrst}
  \{\mathcal{Q}, I\} = \int_\gamma \{\mathcal{Q}, W_k\} = i\int_\gamma dW_{k-1} = 0\;.   
\end{equation}
It proves that $I(\gamma)$ is $\delta$-invariant and, then, a possible observable. To be a global observable of the topological theory, we just have to prove that $I(\gamma)$ is BRST exact, which can be immediately verified taking $\gamma$ as the boundary $\partial\beta$, and applying the Stokes theorem, 
\begin{equation} \label{BrstIexact}
I(\gamma) = \int_{\partial \beta} W_k = \int_\beta dW_k = i\{\mathcal{Q}, \int_\beta W_{k+1}\}\;.
\end{equation}
We conclude, from equations \eqref{Ibrst} and \eqref{BrstIexact},  that $I(\gamma)$ are the global observables of the model as their expectation values produce metric independent quantities, \textit{i.e.}, topological invariants for four-manifolds. Finally, the general path integral representation of Donaldson invariants in Witten's TQFT takes the form
\begin{equation} \label{Donaldsoninv}
Z(\gamma_1, \cdots, \gamma_r) = \int \mathcal{D}\Phi \left( \prod_i \int_{\gamma_i} W_{k_i} \right) e^{-S_W} = \langle \prod_i \int_{\gamma_i} W_{k_i}  \rangle\;, 
\end{equation}
with moduli space dimension
\begin{equation}
\text{dim}(\mathcal{M}) = \sum_i^r(4-k_r)\;.
\end{equation}
One of the beautiful results is the appearing of $W_4$ in the descent equations. Up to a sign, the observable \begin{equation}
    \int_\gamma W_4 = - \frac{1}{2} \int_\gamma F \wedge F
\end{equation}
is the Pontryagin action written in the formalism of p-forms. The Pontryagin action, a well-known topological invariant of four-manifolds, naturally appear as one of the Donaldson polynomials---with a trivial winding number in this case, since $U(W_4) = 0$, and consequently the dimension of the moduli space vanishes. 

\subsection{Baulieu-Singer off-shell approach}\label{BStheory}

Let us now turn to the main properties Baulieu-Singer approach for TQFTs \cite{Baulieu:1988xs}, which is based on an off-shell BRST symmetry, built from the gauge fixing of an original action composed of topological invariants.  

\subsubsection{BRST symmetry in topological gauge theories}

The four-dimensional spacetime is assumed to be Euclidean and flat\footnote{Throughout this work we consider flat Euclidean spacetime. Although the topological action is background independent, the gauge-fixing term entails the introduction of a background. Ultimately, background independence is recovered at the level of correlation functions due to BRST symmetry \cite{Baulieu:1988xs,Blau:1990nv,Abud:1991mu}.}. The non-Abelian topological action $S_0[A]$  in four-dimensional spacetime representing the topological invariants is the Pontryagin action\footnote{It is worth mentioning that the action $S_0[A]$ could encompass a wide range of topological gauge theories. The Pontryagin action is the most common case because it can be defined for all semi-simple Lie groups. Nevertheless, other cases can also be considered. For instance, Gauss-Bonnet and Nieh-Yang topological gravities can be formulated for orthogonal groups \cite{Mardones:1990qc}.},  
\begin{equation} \label{Pontryagin}
S_0[A] = \frac{1}{2}\int d^4x\, F^a_{\mu\nu} \widetilde{F}^a_{\mu\nu}\;,
\end{equation}
that labels topologically inequivalent field configurations, as $S_0[A] = 32 \pi^2 k$, in which $k$ is the topological charge known as winding number. We must note that the Pontryagin action has two different gauge symmetries to be fixed, these are:

(i) the gauge field symmetry,
        \begin{equation}
            \delta A_{\mu}^{a}=D_{\mu}^{ab}\omega^{b}+\alpha_{\mu}^{a}\;;
            \label{eqn:gluon-symmetry}
        \end{equation}

(ii) the topological parameter symmetry,
        \begin{equation}
            \delta\alpha_{\mu}^{a}=D_{\mu}^{ab}\lambda^{b}\;.
            \label{eqn:top-parameter-symmetry}
        \end{equation}
where $D_\mu^{ab} \equiv \delta^{ab}\partial_\mu - gf^{abc}A^c_\mu$ are the components of the covariant derivative in the adjoint representation of the Lie group $G$, $f^{abc}$ are the structure constants of G,  while  $\omega^a$, $\alpha^a_\mu$ and $\lambda^a$ are the infinitesimal $G$-valued gauge parameters. As a consequence of \eqref{eqn:gluon-symmetry}, the field strength also transforms as a gauge field\footnote{The antisymmetrization index notation here employed means that, for a generic tensor, $S_{[\mu\nu]}=S_{\mu\nu}-S_{\nu\mu}$.},
        \begin{equation}
            \delta F_{\mu\nu}^{a}=-gf^{abc}\omega^{b}F_{\mu\nu}^{c}+D_{[\mu}^{ab}\alpha_{\nu]}^{b}\;.
            \label{eqn:F-symmetry}
        \end{equation}
        
The first parameter ($\omega^a$) reflects the usual Yang-Mills symmetry of $S[A]$, whereas the second one ($\alpha^a_\mu$) is the topological shift associated to the fact that $S[A]$ is a topological invariant, \textit{i.e.}, invariant under continuous deformations. The third gauge parameter ($\lambda^a$) is due to an internal ambiguity present in the gauge transformation of the gauge field \eqref{eqn:gluon-symmetry}. The transformation of the gauge field is composed by two independent symmetries. If the space has a boundary, the parameter $\alpha^a_\mu(x)$ must vanish at this boundary but not $\omega^a(x)$, what explains the internal ambiguity described by \eqref{eqn:top-parameter-symmetry} in which $\alpha^a_\mu(x)$ is absorbed into $\omega^a(x)$, and not the other way around \cite{Baulieu:1988xs}. 

Following the BRST quantization procedure, the gauge parameters present in the gauge transformations \eqref{eqn:gluon-symmetry}-\eqref{eqn:F-symmetry} are promoted to ghost fields: $\omega^a \rightarrow c^a$, $\alpha^a_\mu \rightarrow \psi^a_\mu$, and $\lambda^a \rightarrow \phi^a$; $c^a$ is the well-known Faddeev-Popov (FP) ghost; $\psi^a_\mu$ is a topological fermionic ghost; and $\phi^a$ is a bosonic ghost. The corresponding BRST transformations are
\begin{eqnarray}
sA_\mu^a&=&-D_\mu^{ab}c^b+\psi^a_\mu,\nonumber\\
sc^a&=&\frac{g}{2}f^{abc}c^bc^c+\phi^a,\nonumber\\
s\psi_\mu^a&=&gf^{abc}c^b\psi^c_\mu+D_\mu^{ab}\phi^b,\nonumber\\
s\phi^a&=&gf^{abc}c^b\phi^c,\label{brst1}
\end{eqnarray}
from which one can easily check the nilpotency of the BRST operator,
\begin{equation}
s^2 = 0\;,
\end{equation}
by applying two times the BRST operator $s$ on the fields. Naturally, $S_0[A]$ is invariant under the BRST transformations \eqref{brst1}. The nilpotency property of $s$ defines the cohomology of the theory, which allows for the gauge fixing of the Pontryagin action in a BRST invariant fashion. Furthermore, such a property is related to the geometric structure of the off-shell BRST transformations in non-Abelian topological gauge theories.

\subsubsection{Geometric interpretation}\label{geometric}

In order to simplify equations in the following sections, we will employ again the formalism of differential forms. In this formalism, the fields $c$ and $\phi$ are 0-forms, $\psi$ is the 1-form $\psi_\mu dx_\mu$, and $F$, the following 2-form: 
\begin{equation}\label{Fpform}
F = dA + A\wedge A = \frac{1}{2}F_{\mu\nu} dx_\mu \wedge dx_\nu\;,
\end{equation}
where $A$ is the 1-form $A_\mu dx^\mu$, and ``$\wedge$" is the \textit{wedge product} which indicates that the tensor product is completely antisymmetric, and $d$ is the exterior derivative\footnote{The exterior derivative operation in the space of smooth $p$-forms, $\Lambda_p$, $d:\Lambda_p \rightarrow \Lambda_{p+1}$, on a generic $p$-form $\omega_p$, 
\begin{equation}
    \omega_p = \omega_{i_1, i_2, \dots, i_p} dx^{i_1} \wedge dx^{i_2} \cdots \wedge dx^{i_p}\;,
\end{equation}
is locally defined by
\begin{equation}
    d\omega_p = \frac{\partial \omega_{i_1, i_2, \dots, i_p}}{\partial x^j} dx^j \wedge dx^{i_1} \wedge dx^{i_2} \cdots \wedge dx^{i_p}\;,
\end{equation}
being $\omega_p$ a $p$-form, $d\omega_p$ is a $(p+1)$-form. It follows that the exterior derivative is nilpotent, $d^2 = 0$, due to the antisymmetric property of the indices. One assumes that $s$ anticommutes with $d$, $\{s,d\}=0$.}. With this we can then write the BRST transformations in the form 
\begin{eqnarray}\label{BRSTtrans_p-form}
 sA &=& Dc + \psi\;, \nonumber\\
 sc &=& \frac{1}{2}[c,c]+\phi\;, \nonumber\\
 s\psi &=& D\phi + [c, \psi]\;, \nonumber\\
 s\phi &=& [c,\phi]\;.
\end{eqnarray}
The geometric meaning of the topological BRST transformations of \eqref{BRSTtrans_p-form} is revealed from the definition of the extended exterior derivative, $\widetilde{d}$, as the sum of the ordinary exterior derivative with the BRST operator, 
\begin{equation}
    \widetilde{d} = d +s\;,
\end{equation}
and the generalized connection 
\begin{equation} \label{A+c}
    \widetilde{A} = A 
    + c\;.
\end{equation}
By direct inspection one sees that the BRST transformations can be written in terms of the generalized curvature\footnote{The nature of $\phi$ as the ``curvature" in the in instanton moduli space direction is implicit in the BRST transformation of the FP ghost, that can be rewritten in the geometric form $sc + \frac{1}{2} [c,c] = \phi$.}
\begin{equation} \label{univsersalF}
\mathcal{F} = F + \psi +\phi\;,
\end{equation}
such that
\begin{equation} \label{Fcurv}
    \mathcal{F} = \widetilde{d}\widetilde{A} + \frac{1}{2} [\widetilde{A},\widetilde{A}] \;,
\end{equation}
with the Bianchi identity
\begin{equation}\label{FBianchi}
    \widetilde{D}\mathcal{F} = \widetilde{d}\mathcal{F} + [\widetilde{A}, \mathcal{F}] = 0\;.
\end{equation}
Here, the space is graded as a sum of form degree and ghost number, in which the BRST operator is the exterior differential operator in the moduli space direction  $\mathcal{A}/\mathcal{G}$, where the gauge fields that differ by a gauge transformation are identified. The whole space is then $M \times \mathcal{A}/\mathcal{G}$, being $M$ a four-dimensional manifold. According to the gauges worked out in this paper, $M$ will be an Euclidean flat space. 

In the definition \eqref{A+c} and following equations we are adding quantities with different form degrees and ghost numbers as thought they were of the same nature. Obviously this is not being done directly. We must see equations \eqref{Fcurv} and \eqref{FBianchi} as an expansion in form degrees and ghost numbers in which the elements with the same nature on both sides have to be compared. The relevant cohomology is defined by the cohomology of $M \times \mathcal{A}/\mathcal{G}$, ${\widetilde{d}}^2 = 0$, being valid without requiring equations of motion. Such a geometric structure reveals the BRST \textit{off-shell} character of the BS approach\footnote{For a detailed study on the geometric interpretation of the universal fiber bundle and its curvature, we suggest for instance \cite{Daniel:1979ez,Weis:1997kj}.}. We will discuss in Section \ref{observables} how the universal curvature $\mathcal{F}$ generates the same global observables of Witten theory, \textit{i.e.}, the Donaldson polynomials.

\subsubsection{Doublet theorem and gauge fixing: Baulieu-Singer gauges}

Let us recall the \textit{doublet theorem} \cite{Piguet:1995er} which will be indispensable to fix the gauge ambiguities without changing the physical content of the theory. Suppose a theory that contains a pair of fields or sources that form a doublet, \textit{i.e.}, \begin{eqnarray} \label{doublet}
\hat{\delta}\mathcal{X}_i &=& \mathcal{Y}_i \;, \nonumber \\
\hat{\delta} \mathcal{Y}_i &=& 0\;,
\end{eqnarray}
where $i$ is a generic index, and $\hat{\delta}$ is a fermionic nilpotent operator. The field (source) $\mathcal{X}_i$ is assumed to be fermionic. As the operator $\hat{\delta}$ increases the ghost number in one unity by definition, if $\mathcal{X}_i$ is an anti-commuting quantity, $\mathcal{Y}_i$ is a commuting one. The doublet structure of $(\mathcal{X}_i, \mathcal{Y}_i)$ in eq. \eqref{doublet} implies that such fields (or sources) belong to the trivial part of the cohomology of $\hat{\delta}$. The proof is as follows. First, we define the operators 
\begin{eqnarray}
\hat{N} &=& \int dx \left( \mathcal{X}_i \frac{\partial}{\partial \mathcal{X}_i} + \mathcal{Y}_i \frac{\partial}{\partial \mathcal{Y}_i} \right)\;, \\
\hat{A} &=& \int dx \, \mathcal{X}_i \frac{\partial}{\partial \mathcal{Y}_i}\;, \\ 
\hat{\delta} &=& \mathcal{Y}_i \frac{\partial}{\partial \mathcal{X}_i}\;,
\end{eqnarray}
which obey the commutation relations
\begin{eqnarray}
\{ \hat{\delta}, \hat{A} \} &=& \hat{N} \;, \label{algebra1}\\
\left[ \hat{\delta}, \hat{N} \right]&=&0\;, \label{algebra2}
\end{eqnarray}
where $\hat{\delta}$ is a nilpotent operator as it is fermionic, $\hat{\delta}^2=0$. The operator $\hat{N}$ is the counting operator. Being $\bigtriangleup$ a polynomial in the fields, sources and parameters, the cohomology of the nilpotent operator $\hat{\delta}$, as we know, is given by the the solutions of
\begin{equation} \label{deltabig}
\hat{\delta} \bigtriangleup = 0\;,
\end{equation}
that is not exact, \textit{i.e.}, that cannot be written in the form
\begin{equation}
    \bigtriangleup = \hat{\delta} \Sigma\;.
\end{equation}
The general expression of $\bigtriangleup$ is then
\begin{equation}
\bigtriangleup = \widetilde{\bigtriangleup} + \hat{\delta} \Sigma\;,
\end{equation}
where $\widetilde{\bigtriangleup}$ belongs to the non-trivial part of the cohomology, in other words, it is closed, $\hat{\delta}\widetilde{\bigtriangleup} =0$, but not exact, $\widetilde{\bigtriangleup} \neq \hat{\delta} \widetilde{\Sigma}$. One can expand $\bigtriangleup$ in eigenvectors of $\hat{N}$, 
\begin{equation}\label{expansion}
    \bigtriangleup = \sum_{n \geq 0} \bigtriangleup_n\;,
\end{equation}
such that $\hat{N} \bigtriangleup_n = n \bigtriangleup_n$, where $n$ is the total number of $\mathcal{X}_i$ and $\mathcal{Y}_i$ in $\bigtriangleup_n$. Such an expansion is consistent as each $ \bigtriangleup_n$ is a polynomial in $\mathcal{X}_i$ and $\mathcal{Y}_i$, and $\hat{\delta}\bigtriangleup_n=0$ for $\forall n \geq 1$, according to \eqref{doublet} and the commuting properties of $\mathcal{X}_i$ and $\mathcal{Y}_i$. Finally, rewriting \eqref{expansion} as
\begin{equation}
    \bigtriangleup = \bigtriangleup_0 + \sum_{n \geq 1} \frac{1}{n} \hat{N}\bigtriangleup_n\;,
\end{equation}
and then, using the commuting relation \eqref{algebra1}, we get
\begin{equation}
\bigtriangleup = \bigtriangleup_0 + \hat{\delta} \left( \sum_{n \geq 1} \frac{1}{n} \hat{A}\bigtriangleup_n\right)\;,
\end{equation}
which shows that all terms which contain at least one field (source) of the doublet never enter the non-trivial part of the cohomology of $\hat{\delta}$, being thus non-physical---for a more complete analysis, see for instance \cite{Piguet:1995er,Vandersickel:2011zc}.

In order to fix the three gauge symmetries of the non-Abelian topological theory \eqref{eqn:gluon-symmetry}-\eqref{eqn:F-symmetry} we introduce the following three BRST doublets:
\begin{eqnarray} \label{BRSTdoublets}
s\bar{c}^a&=&b^a\;,\;\;\;\;\;\;\;\;sb^a\;=\;0\;,\nonumber\\
s\bar{\chi}^a_{\mu\nu}&=&B_{\mu\nu}^a\;,\;\;sB_{\mu\nu}^a\;=\;0\;,\nonumber\\
s\bar{\phi}^a&=&\bar{\eta}^a\;,\;\;\;\;\;\;\;s\bar{\eta}^a\;=\;0\;,\label{brst2}
\end{eqnarray}
where $\bar{\chi}^a_{\mu\nu}$ and $B_{\mu\nu}^a$ are (anti-)self-dual fields according to the (negative) positive sign, see \eqref{gf3BS} below. The $\mathcal{G}$-valued Lagrange multiplier fields $b^a$, $B^a_{\mu\nu}$ and $\bar{\eta}$ have respectively ghost numbers $0$, $0$, and $-1$; while the antighost fields $\bar{c}^a$, $\bar{\chi}^a_{\mu\nu}$ and $\bar{\phi}^a$, ghost numbers $-1$, $-1$ and $-2$. (For completeness and further use, the quantum numbers of all fields are displayed in Table \ref{table1}.)

\begin{table}[h]
\centering
\setlength{\extrarowheight}{.5ex}
\begin{tabular}{cc@{\hspace{-.3em}}cccccccccc}
\cline{1-1} \cline{3-12}
\multicolumn{1}{|c|}{Field} & & \multicolumn{1}{|c|}{$A$} & \multicolumn{1}{c|}{$\psi$} & \multicolumn{1}{c|}{$c$} & \multicolumn{1}{c|}{$\phi$} & \multicolumn{1}{c|}{$\bar{c}$} & \multicolumn{1}{c|}{$b$} &\multicolumn{1}{c|}{$\bar{\phi}$} & \multicolumn{1}{c|}{$\bar{\eta}$} & \multicolumn{1}{c|}{$\bar{\chi}$} & \multicolumn{1}{c|}{$B$}   
\\ \cline{1-1} \cline{3-12} 
\\[-1.18em]
\cline{1-1} \cline{3-12}
\multicolumn{1}{|c|}{Dim} & & \multicolumn{1}{|c|}{1} & \multicolumn{1}{c|}{1} & \multicolumn{1}{c|}{0} & \multicolumn{1}{c|}{0} & \multicolumn{1}{c|}{2} & \multicolumn{1}{c|}{2} &\multicolumn{1}{c|}{2} & \multicolumn{1}{c|}{2} & \multicolumn{1}{c|}{2} & \multicolumn{1}{c|}{2}   
\\
\multicolumn{1}{|c|}{Ghost n$^o$} & & \multicolumn{1}{|c|}{0} & \multicolumn{1}{c|}{1} & \multicolumn{1}{c|}{1} & \multicolumn{1}{c|}{2} & \multicolumn{1}{c|}{-1} & \multicolumn{1}{c|}{0} &\multicolumn{1}{c|}{-2} & \multicolumn{1}{c|}{-1} & \multicolumn{1}{c|}{-1} & \multicolumn{1}{c|}{0}   
\\ \cline{1-1} \cline{3-12} 
\end{tabular}
\caption{Quantum numbers of the fields.}
\label{table1}
\end{table}
Working in Baulieu-Singer gauges amounts to considering the constraints \cite{Baulieu:1988xs}
\begin{eqnarray}
    \label{eqn:gauge-fixingsBS}
    \partial_{\mu} A_{\mu}^{a} &=&- \frac{1}{2}\rho_1 b^a\;, \label{gf1BS}\\
    D_{\mu}^{ab} \psi_{\mu}^{a} &=& 0\;,\label{gf2BS} \\
    F_{\mu\nu}^{a} \pm \widetilde{F}_{\mu\nu}^{a} &=&- \frac{1}{2} \rho_2 B^a_{\mu\nu}\;,\label{gf3BS}
\end{eqnarray}
where $\rho_1$ and $\rho_2$ are real gauge parameters. In a few words, beyond the gauge fixing of the topological ghost \eqref{gf2BS}, we must interpret the requirement of two extra gauge fixings due to the fact that the gauge field possesses two independent gauge symmetries. In this sense, condition \eqref{gf1BS} fixes the usual Yang-Mills symmetry $\delta A^a_\mu = D_\mu^{ab}\omega^b$, and the second one, \eqref{gf3BS}, the topological shift $\delta A^a_\mu = \alpha^a_\mu$. The (anti-)self-dual condition for the field strength (in the limit $\rho_2 \rightarrow 0$) is convenient to identify the well-known observables of topological theories for four-manifolds (see \cite{vanBaal:1989aw}) given by the Donaldson invariants \cite{DONALDSON1990257, Donaldson:1983wm}, that are described in terms of the instantons. 

The partition functional of the topological action in BS gauges \eqref{eqn:gauge-fixingsBS} takes the form 
\begin{equation} \label{ZBS}
   Z_{BS} = \int \mathcal{D}c\mathcal{D}\bar{c}\mathcal{D}\psi_\mu\mathcal{D}\bar{\chi}_{\mu\nu}\mathcal{D}B_{\mu\nu}\mathcal{D}\phi\mathcal{D}\bar{\phi}\mathcal{D}\eta e^{-S_{BS}}\;,
\end{equation}
whereby 
\begin{equation} \label{BSactionfull}
    S_{BS} = S_0[A] + S_{gf}^{BS}\;,
\end{equation}
being $S_{gf}^{BS}$ the gauge-fixing action which belongs to trivial part of the cohomology, given by 
\begin{eqnarray}\label{SgfBS}
 S_{gf}^{BS} &=& s\, \text{Tr}\int d^4x \left[ \bar{\chi}_{\mu\nu}\left( F_{\mu\nu} \pm \widetilde{F}_{\mu\nu} + \frac{1}{2}\rho_2 B_{\mu\nu}\right) + \bar{\phi} D_\mu \psi_\mu + \bar{c}\left(\partial_\mu A_\mu - \frac{1}{2}\rho_1 b \right)\right]\nonumber\\ 
 &=& \text{Tr}\int d^4x \left[ B_{\mu\nu}\left( F_{\mu\nu} \pm \widetilde{F}_{\mu\nu} + \frac{1}{2}\rho_2 B_{\mu\nu}\right) + \bar{\chi}_{\mu\nu}\left( D_{[\mu}\psi_{\nu]} \pm \frac{1}{2} \varepsilon_{\mu\nu\alpha\beta} D_{[\alpha}\psi_{\beta]}\right)\right.\nonumber\\
 &-&\left.\bar{\chi}_{\mu\nu} \left[c, F_{\mu\nu} \pm \widetilde{F}_{\mu\nu}\right]+\eta D_\mu \psi_\mu + \bar{\phi}\left[\psi_\mu, \psi_\mu\right] + \bar{\phi} D_\mu D_\mu \phi - b\left(\partial_\mu A_\mu - \frac{1}{2}\rho_1b \right)\right.\nonumber\\
 &-&\left.\bar{c} \partial_\mu D_\mu c -\bar{c}\partial_\mu \psi_\mu \right] \;.
\end{eqnarray}

A key observation is that, for $\rho_1 =\rho_2 =1$, one can eliminate the topological term $S_0[A]$, \textit{i.e.}, the Pontryagin action, by integrating out the field $B_{\mu\nu}$, such that
\begin{equation}
\text{Tr}\{B_{\mu\nu}\left(F_{\mu\nu} + \widetilde{F}_{\mu\nu}\right) + \frac{1}{2}B_{\mu\nu}B_{\mu\nu}\} \rightarrow \text{Tr}\{F_{\mu\nu}F_{\mu\nu} + F_{\mu\nu}\widetilde{F}_{\mu\nu}\} \;.
\end{equation}
In this case we obtain a classical topological action which is equivalent to a Yang-Mills action plus ghost interactions. Such an action, however, does not produce local observables as the cohomology of the theory remain the same, as we will discuss in more detail later in Section \ref{observables}. 

Another important property is that the Green functions of local observables in \eqref{ZBS} do not depend on the choice of the background metric \cite{Baulieu:1988xs}. Let $S_{BS}^g$ be an action with metric choice $g_{\mu\nu}$, and $S_{BS}^{g+\delta g}$, the same action up to the change of $g_{\mu\nu}$ into $g_{\mu\nu} + \delta g_{\mu\nu}$. As the only terms depending on the metric belong to the trivial part of cohomology we conclude immediately that $S_{BS}^g$ and $S_{BS}^{g+\delta g}$ only differ by a BRST-exact term, 
\begin{equation}
    S_{BS}^g - S_{BS}^{g+\delta g} = s \int d^4x \bigtriangleup^{(-1)} \;,
\end{equation}
where $\bigtriangleup^{(-1)}$ is a polynomial of the fields, with ghost number $-1$. It means that the expectation values of local operators are the same if computed with a background metric $g_{\mu\nu}$ or $g_{\mu\nu} + \delta g_{\mu\nu}$, 
\begin{equation} \label{metricindBS}
\frac{\delta}{\delta g_{\mu\nu}} \langle \prod_p \mathcal{O}_{\alpha_p}(\phi_i) \rangle = 0\;,
\end{equation}
where $\mathcal{O}_{\alpha_p}(\phi_i)$ are functional operators of the quantum fields $\phi_i(x)$, see eq. \eqref{toporesult}. An anomaly in the topological BRST symmetry would break the above equation. However there is no $4$-form with ghost number 1, $\bigtriangleup^{(1)}_{4-\text{form}}$, defined modulo $s$- and $d$- exact terms which obeys (cf. \cite{Baulieu:1988xs})
\begin{equation} \label{anomalycondition}
s\bigtriangleup^{(1)}_{4-\text{form}} + d \bigtriangleup^{(2)}_{3-\text{form}} = 0\;.
\end{equation}
Therefore, radiative corrections which could break the topological property \eqref{metricindBS} at the quantum level are not expected. The formal proof of the absence of gauge anomalies to all orders in the topological BS theory is achieved by employing the isomorphism described in  \cite{Dixon:1979bs,Delduc:1996yh}.

\subsubsection{Absence of gauge anomalies}

The proof of the absence of gauge anomalies for the Slavnov-Taylor identity, 
\begin{equation} \label{STid}
\mathcal{S}(S) = 0\;,
\end{equation}
consists in proving that the cohomology of $\mathcal{S}$ is empty. In the equation above, $S$ is the classical action for a given gauge choice, and
\begin{equation} \label{STid2}
\mathcal{S} = \int d^4x\, (s\Phi^I) \frac{\delta}{\delta \Phi^I}\;,
\end{equation}
where $\Phi^I$ represents all fields. As $\mathcal{S}$ is a Ward identity, in the absence of anomalies the symmetry \eqref{STid} is also valid at the quantum level, \textit{i.e.}, $\mathcal{S}(\Gamma)=0$, being $\Gamma$ the quantum action in loop expansion. In eq. \eqref{STid2}, $s\Phi^I$ represents the BRST transformation of each field $\Phi^I$. The fields $\bar{c}$, $b$, $\bar{\chi}_{\mu\nu}$, $B_{\mu\nu}$, $\bar{\phi}$ and $\bar{\eta}$ transform as doublets, cf. eq. \eqref{BRSTdoublets}. Changing the variables according to the redefinitions
\begin{eqnarray} \label{redefinitions}
    \psi &\rightarrow & \psi^\prime= \psi - Dc\;, \nonumber \\
    \phi &\rightarrow & \phi^\prime = \phi - \frac{1}{2} [c,c]\;,
\end{eqnarray}
the BRST transformations \eqref{BRSTtrans_p-form} are reduced to the doublet transformations
\begin{eqnarray}
    s A &=& \psi^\prime\;, \nonumber \\ 
    s\psi^\prime &=& 0\;,\nonumber \\ 
    sc &=& \phi^\prime\;,\nonumber \\
    s\phi^\prime &=& 0\;.\nonumber \\
\end{eqnarray}
It configures a reduced transformation in which the non-linear part of the  BRST transformations in the Slavnov-Taylor identity were eliminated. The complete transformation in this space is given by the reduced operator
\begin{equation} \label{STdoublet}
\mathcal{S}_{doublet} = \int d^4x\, (s{\Phi^\prime}^I) \frac{\delta}{\delta {\Phi^\prime}^I}\;,
\end{equation}
where ${\Phi^\prime} = \{A, \psi^\prime, c, \phi^\prime, \bar{c}, b, \bar{\chi}_{\mu\nu}, B_{\mu\nu}, \bar{\phi}, \eta\}$, which is composed of five doublets. It means that $\mathcal{S}_{doublet}$ has vanishing cohomology ($H$), 
\begin{equation} \label{subspaceHzero}
H(\mathcal{S}_{doublet}) = \varnothing\;,
\end{equation}
in other words, that any polynomial of the fields $\Phi^\prime$, $\bigtriangleup(\Phi^\prime)$, that satisfies
\begin{equation}
\mathcal{S}_{doublet}(\bigtriangleup(\Phi^\prime)) = 0\;,
\end{equation}
belonging to the trivial part of the cohomology of $\mathcal{S}_{doublet}$ (see the doublet theorem in previous section). The crucial point here is the fact that the cohomology of $\mathcal{S}$ in the
space of local integrated functionals in the fields and sources is isomorphic to a subspace of $H(\mathcal{S}_{doublet})$. Consequently $\mathcal{S}$ has also vanishing cohomology  \cite{Dixon:1979bs,Sorella:1989ri,Delduc:1996yh}, \begin{equation} \label{cohomology}
    H(\mathcal{S}) = \varnothing\;.
\end{equation}
The result \eqref{cohomology} shows that there is no room for an anomaly in the Salvnov-Taylor identity \eqref{STid}. All counterterms at the quantum level will belong to the trivial part of cohomology, and the condition \eqref{anomalycondition} for the existence of an anomaly capable of breaking the topological property \eqref{metricindBS} never occurs. Due to the algebraic structure of the theory, eq. \eqref{cohomology} proves that all Ward identities are free of gauge anomalies, cf. \cite{Delduc:1996yh}. As a consequence of this result, the background metric independence is valid to all orders in perturbation theory.

The second point, and not least, is the conclusion that the BS theory has no local observables. Due to its vanishing cohomology \eqref{cohomology}, all BRST-invariant quantities must belong to the non-physical (or trivial) part of the cohomology of $s$, and the only possible observables are the global ones, \textit{i.e.}, topological invariants for four-manifolds. Such observables are characterized by the cohomology of $s$ \cite{Ouvry:1988mm, Sorella:1989ri}, in which the observables are globally defined in agreement with the supersymmetric formulation of J. H. Horne \cite{Horne:1988yn}. A simple way to identify theses observables is accomplished by studying the cohomology of the extended space $M \times \mathcal{A}/ \mathcal{G}$, where the metric independent observables, known as \textit{Chern classes}, are constructed in terms of the universal curvature $\mathcal{F}$ \eqref{univsersalF}. The Donaldson polynomials are naturally recovered, characterized by the so-called \textit{equivariant cohomology}, that relates the BS approach to Witten theory at the level of observables.

\subsubsection{Equivariant cohomology and global observables}\label{observables}

Witten's topological theory is constructed  without fixing its remaining ordinary Yang-Mills gauge symmetry. The theory is developed in the instanton moduli space $\mathcal{A}/\mathcal{G}$. A generic  observable of his theory, $\mathcal{O}^{(W)}_{\alpha_i}$, is naturally gauge invariant under Yang-Mills gauge transformations, 
\begin{equation}
s_{YM} \mathcal{O}^{(W)}_{\alpha_i}= 0\;,
\end{equation}
where $s_{YM}$ is the nilpotent BRST operator related to the ordinary Yang-Mills symmetry, \textit{i.e.}, without including the topological shift, namely,
\begin{eqnarray}
    s_{YM} A_\mu &=& D_\mu c\;, \nonumber \\
    s_{YM} \Phi_{adj} &=&  [c, \Phi_{adj}]\;,
\end{eqnarray}
where $\Phi_{adj}$ is a generic field in adjoint representation. We conclude that we can add an ordinary Yang-Mills gauge transformation (in the $\mathcal{A}/\mathcal{G}$ direction) to Witten fermionic symmetry based on the ``topological shift" $\delta A_\mu \sim \psi_\mu$,
\begin{equation}
    \delta \rightarrow  \delta_{eq} = \delta  + s_{YM}\;,
\end{equation}
in such a way that the descent equations for $\delta \sim \{\mathcal{Q}, \,\cdot\,\}$ will remain the same, see \eqref{deltaQ} and \eqref{deltadescent1}-\eqref{deltadescentf}. The operator $\delta_{eq}$ is nilpotent when acting on gauge-invariant quantities under YM transformations, defining thus a cohomology in a space where the fields that differ by a Yang-Mills gauge transformations are identified, known as \textit{equivariant cohomology}. Such a property indicates that there is a link between Witten theory and BS approach in which the BRST operator, $s$, is naturally defined taking into account the topological shift and the ordinary Yang-Mills transformation in a single formalism.

To prove the link between both approaches, we must remember that the universal curvature in the space $M \times \mathcal{A}/\mathcal{G}$ is given by the sum $\mathcal{F}=F+\psi+\phi$. The difference between the on-shell BRST operator, $s$, and the Witten fermionic symmetry, $\delta$, for $\mathbb{X} = (F, \psi, \phi)$ is of the form
\begin{equation}
    s \mathbb{X} = \delta \mathbb{X} + [\mathbb{X}, c]\;, 
\end{equation}
in other words, in the space of the fields $(F, \psi, \phi)$, $s$ and $\delta$ differ by an ordinary Yang-Mills transformation, as $(F, \psi, \phi)$ transform in the adjoint representation of the gauge group. These fields are the only ones we need to obtain the Donaldson polynomials as the observables of the BS theory, since in the space $M \times \mathcal{A}/\mathcal{G}$ they are constructed in terms of $\mathcal{F}$. This allows for identifying the equivariant operator with the BRST one, $\delta_{eq} \equiv s$, according to the construction of the observables in Witten and BS theory, respectively.

To understand the above statement, we must invoke the $n$'th Chern class, $\widetilde{\mathcal{W}}_n$, defined in terms of the universal curvature by
\begin{equation}\label{Wn}
   \widetilde{\mathcal{W}}_n = \text{Tr}\,(\underbrace{ \mathcal{F} \wedge \mathcal{F} \wedge \cdots \wedge \mathcal{F}}_{\mbox{n times}})
\end{equation}
where $n= \{1,2,3, \cdots\}$ is the number of wedge products. $\widetilde{\mathcal{W}}_n$ represents the most general observables of BS theory\footnote{It is not possible to construct topological observables using the Hodge product, as it is metric dependent. For this reason we never obtain Yang-Mills terms of the type $\{\text{Tr}(F_{\mu\nu}F^{\mu\nu}), \text{Tr}(F_{\mu\nu}F^{\nu\sigma}F^{\mu}\;_\sigma), \cdots\}$, without Levi-Civita tensors in the internal product, in the place of metric tensors. Moreover, the Wilson loop
\begin{equation}
    W_{P}^{(C)} = \text{Tr}\{\mathcal{P} e^ {i\oint_C A_\mu dx_\mu}\}\;,
\end{equation}
is not an observable in the non-Abelian topological BS case, as it is not gauge-invariant due to the topological shift symmetry. In any case, it does not make sense to discuss confinement in the BS theory, as it is not confining to any energy scale. Thence, the only possibilities for topological invariants are the wedge products in $\widetilde{\mathcal{W}}_n$.}. Weyl theorem \cite{Weis:1997kj} ensures that $\widetilde{\mathcal{W}}_n$ is closed with respect to the extended differential operator $\widetilde{d}= d +s$ \cite{Baulieu:1988xs, Kanno:1988wm}, \textit{i.e.}, 
\begin{equation} \label{dW=0}
    \widetilde{d}\, \widetilde{\mathcal{W}}_n = 0\;.
\end{equation}
If we choose the first Chern class
\begin{equation}
\widetilde{\mathcal{W}}_1 = \text{Tr} \, (\mathcal{F} \wedge \mathcal{F})\;,     
\end{equation}
the expansion in ghost numbers of equation \eqref{dW=0} yields
\begin{eqnarray}
s \text{Tr}\, (F \wedge F) &=& d \text{Tr}\,(    -2\psi \wedge F)\;, \\
s \text{Tr}\, (\psi \wedge F) &=& d \text{Tr}\,(    -\frac{1}{2}\psi \wedge \psi - \phi F)\;, \\
s \text{Tr}\, (\psi \wedge \psi +2\phi F) &=& d \text{Tr}\,(    2\psi\phi)\;, \\
s \text{Tr}\, (\psi\phi) &=& d \text{Tr}\,(    -\frac{1}{2}\phi\phi)\;, \\
s \text{Tr}\, (\phi\phi) &=& 0\;,
\end{eqnarray}
which are the same descent equations obtained in \eqref{deltadescent1}-\eqref{deltadescentf} following Witten analysis, only replacing $\delta$ (or $\delta_{eq}$) by $s$, proving that Baulieu-Singer and Witten topological theories possess the same observables given by the Donaldson invariants \eqref{Donaldsoninv}.

It should not seem surprising the fact that the observables in the BS approach are naturally invariant under ordinary Yang-Mills symmetry, as the $n$'th Chern class is Yang-Mills invariant itself \eqref{Wn} since $\mathcal{F}$ transforms in the adjoint representation of the gauge group. Equation \eqref{dW=0} provides a powerful tool to obtain Donaldson polynomials for any ghost number. One must note that we do not have to worry about with the independence of Faddeev-Popov ghosts to construct the observables in the BS approach. Although the gauge-fixed BS action has FP ghosts due to the gauge fixing of the Yang-Mills ambiguity, the $(c, \bar{c})$ independence of $\widetilde{\mathcal{W}}_n$ is a direct consequence of the fact that the universal curvature of the space $M \times \mathcal{A}/\mathcal{G}$ does not depend on FP ghosts, but only on $F$, and the ghosts $\psi$ and $\phi$.

In the weak coupling limit of the twisted $N=2$, the observables of both theories are undoubtedly the same: the topological Donaldson invariants \cite{Weis:1997kj, Delduc:1996yh, Boldo:2003jq}. We might ask if the quantum behavior are also compatible, once BS and Witten actions does not differ by a BRST-exact term, 
\begin{equation}\label{BSWittenrelation}
S_{BS} - S_W = \Sigma_\mathcal{G} \neq s(\cdots)\;.    
\end{equation}
The relation above does not depend on the gauge choice. Consequently, we cannot say, in principle, that BS and Witten partition functions are equivalent at quantum level, since
\begin{equation}
Z_{BS} = \int \mathcal{D}\Phi e^{-S_{BS}} = \int \mathcal{D}\Phi e^{-S_{W}-\Sigma_{\mathcal{G}}}\;,
\end{equation}
wherein $\Sigma_{\mathcal{G}}$ is not $s$-exact. At a first view, $Z_{BS} \neq Z_W = \int \mathcal{D}\Phi e^{-S_{W}}$. The fact that $\Sigma_{\mathcal{G}} \neq s (\cdots)$ opens the possibility for both theories to have different quantum properties. The one-loop exactness of twisted $N=2$ SYM $\beta$-function for instance is a well-known result in literature \cite{Blasi:2000qw}. We will now analyze the Ward identities of the BS theory in self-dual Landau gauges, in order to compare the quantum properties of the DW and BS theories.

\section{Quantum properties of BS theory in the self-dual Landau gauges}\label{SDLgauges}

In this section we will summarize the quantum properties of BS theory in the self-dual Landau (SDL) gauges\footnote{For simplicity, throughout the text we will refer to the Baulieu-Singer theory in the self-dual Landau gauges as self-dual BS theory.}. Extra details can be found in \cite{Brandhuber:1994uf,WerneckdeOliveira:1993pa,Junqueira:2017zea,Junqueira:2018xgl, Junqueira:2018zxr}.

\subsection{Absence of radiative corrections}\label{Absence}

Working in the self-dual Landau gauges amounts to considering the constraints \cite{Brandhuber:1994uf}
\begin{eqnarray}
    \label{eqn:gauge-fixings}
    \partial_{\mu} A_{\mu}^{a} &=& 0\;, \label{gf1}\\
    \partial_{\mu} \psi_{\mu}^{a} &=& 0\;,\label{gf2} \\
    F_{\mu\nu}^{a} \pm \widetilde{F}_{\mu\nu}^{a} &=& 0\;.\label{gf3}
\end{eqnarray}
Through the introduction of the three BRST doublets described in eq. \eqref{BRSTdoublets}, the complete gauge-fixed topological action in SDL gauges takes the form
\begin{equation}\label{Stotal}
S[\Phi] = S_0[A] + S_{gf}[\Phi]\;,
\end{equation}
with $S_0[A]$ standing for the Pontryagin action, and
\begin{eqnarray}\label{S_gf}
S_{gf}\left[\Phi\right]&=&s\int d^4z\left[\bar{c}^a\partial_\mu A_\mu^a+\frac{1}{2}\bar{\chi}^a_{\mu\nu}\left(F_{\mu\nu}^a\pm\widetilde{F}_{\mu\nu}^a\right)+\bar{\phi}^a\partial_\mu\psi^a_\mu\right]\nonumber\\
&=&\int d^4z\left[b^a\partial_\mu A_\mu^a+\frac{1}{2}B^a_{\mu\nu}\left(F_{\mu\nu}^a\pm\widetilde{F}_{\mu\nu}^a\right)+
\left(\bar{\eta}^a-\bar{c}^a\right)\partial_\mu\psi^a_\mu+\bar{c}^a\partial_\mu D_\mu^{ab}c^b+\right.\nonumber\\
&-&\left.\frac{1}{2}gf^{abc}\bar{\chi}^a_{\mu\nu}c^b\left(F_{\mu\nu}^c\pm\widetilde{F}_{\mu\nu}^c\right)-\bar{\chi}^a_{\mu\nu}\left(\delta_{\mu\alpha}\delta_{\nu\beta}\pm\frac{1}{2}\epsilon_{\mu\nu\alpha\beta}\right)D_\alpha^{ab}\psi_\beta^b+\bar{\phi}^a\partial_\mu D_\mu^{ab}\phi^b+\right.\nonumber\\
&+&\left.gf^{abc}\bar{\phi}^a\partial_\mu\left(c^b\psi^c_\mu\right)\right]\;.\label{gfaction}
\end{eqnarray}
This action possesses a rich set of symmetries, see \cite{Junqueira:2017zea} and Appendix \ref{Ap1}. In order to control the non-linearity of the Slavnov-Taylor identity (Equation \eqref{st1}) and the bosonic symmetry $\mathcal{T}$ (Equation \eqref{cnl1}), we have to introduce external sources given by the following three BRST doublets \cite{Brandhuber:1994uf}
\begin{eqnarray} \label{sources1}
s\tau^a_\mu&=&\Omega_\mu^a\;,\;\;\;\;\;\;\;\;s\Omega_\mu^a\;=\;0\;,\nonumber\\
sE^a&=&L^a\;,\;\;\;\;\;\;\;\,\;sL^a\;=\;0\;,\nonumber\\
s \Lambda^a_{\mu\nu} &=&  K^a_{\mu\nu} \;, \quad \,\,\,\,\,sK^a_{\mu\nu}=0\;.
\end{eqnarray}
The respective external action, written as a BRST-exact contribution preserving the the physical content of theory, takes the form
\begin{eqnarray}
S_{ext}&=&s\int d^4z\left(\tau_\mu^aD_\mu^{ab}c^b+\frac{g}{2}f^{abc}E^ac^bc^c+gf^{abc}\Lambda^a_{\mu\nu}c^b\bar{\chi}^c_{\mu\nu}\right)\nonumber\\
&=&\int d^4z\left[\Omega_\mu^aD_\mu^{ab}c^b+\frac{g}{2}f^{abc}L^ac^bc^c+gf^{abc}K^a_{\mu\nu}c^b\bar{\chi}^c_{\mu\nu}+\tau^a_\mu\left(D_\mu^{ab}\phi^b+gf^{abc}c^b\psi_\mu^c\right)\right.\nonumber\\
&+&\left.gf^{abc}E^ac^b\phi^c+gf^{abc}\Lambda^a_{\mu\nu}c^bB^c_{\mu\nu}-gf^{abc}\Lambda^a_{\mu\nu}\phi^b\bar{\chi}^c_{\mu\nu}\right.\nonumber\\
&-&\left.\frac{g^2}{2}f^{abc}f^{bde}\Lambda^a_{\mu\nu}\bar{\chi}^c_{\mu\nu}c^dc^e\right]\;,\label{extaction}
\end{eqnarray}
with the corresponding quantum number of the external sources displayed in Table \ref{table2} below. Therefore, the full classical action to be quantized is 
\begin{equation}
\Sigma[\Phi]=S_0[A]+S_{gf}[\Phi]+S_{ext}[\Phi]\;.\label{fullaction}
\end{equation}
The introduction of the external action does not break the original symmetries, and the physical limit is obtained by setting the external sources to zero \cite{Piguet:1995er}.

\begin{table}[h]
\centering
\setlength{\extrarowheight}{.5ex}
\begin{tabular}{cc@{\hspace{-.3em}}cccccc}
\cline{1-1} \cline{3-8}
\multicolumn{1}{|c|}{Source} & & \multicolumn{1}{|c|}{$\tau$} & \multicolumn{1}{c|}{$\Omega$} & \multicolumn{1}{c|}{$E$} & \multicolumn{1}{c|}{$L$} & \multicolumn{1}{c|}{$\Lambda$} & \multicolumn{1}{c|}{$K$}
\\ \cline{1-1} \cline{3-8} 
\\[-1.18em]
\cline{1-1} \cline{3-8}
\multicolumn{1}{|c|}{Dim} & & \multicolumn{1}{|c|}{3} & \multicolumn{1}{c|}{3} & \multicolumn{1}{c|}{4} & \multicolumn{1}{c|}{4} & \multicolumn{1}{c|}{2} & \multicolumn{1}{c|}{2} \\
\multicolumn{1}{|c|}{Ghost n$^o$} & & \multicolumn{1}{|c|}{-2} & \multicolumn{1}{c|}{-1} & \multicolumn{1}{c|}{-3} & \multicolumn{1}{c|}{-2} & \multicolumn{1}{c|}{-1} & \multicolumn{1}{c|}{0}
\\ \cline{1-1} \cline{3-8}
\end{tabular}
\caption{Quantum numbers of the external sources.}
\label{table2}
\end{table}

One of the symmetries are of particular interest to us: the vector supersymmetry described by eq. \eqref{w2}, cf. \cite{Brandhuber:1994uf, Junqueira:2017zea}. By applying BRST-algebraic renormalization techniques \cite{Piguet:1995er}, and disregarding Gribov ambiguities, it was proved in \cite{Junqueira:2017zea}, with the help of Feynman diagrams, that all two-point functions are tree-level exact, as a consequence of the Ward identities of the model. In particular, as a consequence of the vector supersymmetry \eqref{w1}, the gauge field propagator vanishes to all orders in perturbation theory, 
\begin{eqnarray}
\langle A^a_\mu(p) A^b_\nu(q) \rangle = 0\;.\label{aa}
\end{eqnarray}
In \cite{Junqueira:2018xgl} this result was generalized: not only the two-point functions of the self-dual BS theory are tree-level exact, but all $n$-point Green functions of the model do not receive any radiative corrections. This is a direct consequence of the null gauge propagator \eqref{aa} together with the vertex structure of the full action \eqref{fullaction}. Following the Feynman rules notation of \cite{Junqueira:2018xgl}, we represent the relevant propagators by 
\begin{align}
	\langle AA \rangle &=
	\begin{tikzpicture}
		\begin{feynhand}
			\vertex (1);
			\vertex (2) [right=30pt of 1] ;
			\propag[glu] (1) to (2);
		\end{feynhand}
	\end{tikzpicture}
	\quad ,&
	\langle c\bar{c} \rangle &=
	\begin{tikzpicture}
		\begin{feynhand}
			\vertex (1);
			\vertex (2) [right=30pt of 1];
			\propag[gho] (1) to (2);
		\end{feynhand}
	\end{tikzpicture}
	\quad ,&
	\langle \bar{\chi}\psi \rangle &=
	\begin{tikzpicture}
		\begin{feynhand}
			\vertex (1);
			\vertex (2) [right=30pt of 1];
			\propag[pho] (1) to (2);
		\end{feynhand}
	\end{tikzpicture}
	\quad ,\qquad \nonumber\\
	\langle AB \rangle &=
	\begin{tikzpicture}
		\begin{feynhand}
			\vertex (1);
			\vertex (2)  [right=18pt of 1];
			\vertex (3) [right=12pt of 2];
			\vertex (2^) [above=2pt of 2];
			\vertex (3^) [above=2pt of 3];
			\graph{(1) --[glu] (2) --[plain] (3)};
			\propag[plain] (2^) to (3^);
		\end{feynhand}
	\end{tikzpicture}
	\quad ,&
	\langle	\phi \bar{\phi} \rangle &=
	\begin{tikzpicture}
		\begin{feynhand}
			\vertex (1);
			\vertex (1^) [above=2pt of 1];
			\vertex (2) [right=30pt of 1];
			\vertex (2^) [above=2pt of 2];
			\propag[gho] (1) to (2);
			\propag[gho] (1^) to (2^);
		\end{feynhand}
	\end{tikzpicture}
	\quad .
\label{eq.propags}
\end{align}
The relevant vertices are represented by:
\begin{align}
	&
	\begin{tikzpicture}
		\begin{feynhand}
			\vertex[dot] (0) {};
			\vertex (1) [below right=0pt of 0];
			\vertex (1^) [above=2pt of 1];
			\vertex (1<) [left=25pt of 1];
			\vertex (1^<) [left=25pt of 1^];
			\vertex [above left=0pt of 1<];
			\vertex (2) [above right=25pt of 1];
			\vertex (3) [below right=25pt of 1];
			\propag[plain] (1) to [edge label=$B$] (1<);
			\propag[plain] (1^) to (1^<);
			\propag[glu] (1) to [edge label=$A$](2);
			\propag[glu] (1) to [edge label=$A$](3);
		\end{feynhand}
	\end{tikzpicture}
	\quad ,&&
	\begin{tikzpicture}
		\begin{feynhand}
			\vertex[dot] (1) {};
			\vertex (2) [left=25pt of 1] {};
			\vertex (3) [above right=25pt of 1];
			\vertex (4) [below right=25pt of 1];
			\propag[pho] (1) to [edge label=$\bar\chi$] (2);
			\propag[glu] (1) to [edge label=$A$] (3);
			\propag[gho] (1) to [edge label=$c$] (4);
		\end{feynhand}
	\end{tikzpicture}
	\quad ,&&
	\begin{tikzpicture}
		\begin{feynhand}
			\vertex[dot] (1) {};
			\vertex (2) [left=25pt of 1];
			\vertex (3) [above right=25pt of 1];
			\vertex (4) [below right=25pt of 1];
			\propag[glu] (1) to [edge label=$A$] (2);
			\propag[pho] (1) to [edge label=$\psi$] (3);
			\propag[pho] (1) to [edge label=$\bar\chi$] (4);
		\end{feynhand}
	\end{tikzpicture}
	\quad ,&&
	\begin{tikzpicture}
		\begin{feynhand}
			\vertex[dot] (1) {};
			\vertex (2) [left=25pt of 1];
			\vertex (3) [above right=25pt of 1];
			\vertex (4) [below right=25pt of 1];
			\propag[glu] (1) to [edge label=$A$] (2);
			\propag[gho] (1) to [edge label=$c$] (3);
			\propag[gho] (1) to [edge label=$\bar{c}$] (4);
		\end{feynhand}
	\end{tikzpicture}
	\quad , \qquad \nonumber\\
	&
	\begin{tikzpicture}
		\begin{feynhand}
			\vertex[dot] (1) {};
			\vertex (2) [left=25pt of 1];
			\vertex (3) [above right=25pt of 1];
			\vertex (4) [below right=25pt of 1];
			\vertex (1>) [right=1pt of 1];
			\vertex (1>^) [above right=25pt of 1>];
			\vertex (1>v) [below right=25pt of 1>];
			\propag[glu] (1) to [edge label=$A$] (2);
			\propag[gho] (1) to [edge label=$\bar\phi$] (3);
			\propag[gho] (1) to [edge label=$\phi$] (4);
			\propag[gho] (1>) to (1>^);
			\propag[gho] (1>) to (1>v);
		\end{feynhand}
	\end{tikzpicture}
	\quad ,&&
	\begin{tikzpicture}
		\begin{feynhand}
			\vertex[dot] (0) {};
			\vertex (1);
			\vertex (1^) [above=2pt of 1];
			\vertex (1<) [left=25pt of 1];
			\vertex (1^<) [left=25pt of 1^];
			\vertex [above=0pt of 1^<];
			\vertex (2) [above right=25pt of 1];
			\vertex (3) [below right=25pt of 1];
			\propag[gho] (1) to [edge label=$\bar\phi$] (1<);
			\propag[gho] (1^) to (1^<);
			\propag[pho] (1) to [edge label=$\psi$] (2);
			\propag[gho] (1) to [edge label=$c$] (3);
		\end{feynhand}
	\end{tikzpicture}
	\quad ,&&
	\begin{tikzpicture}
		\begin{feynhand}
			\vertex[dot] (1) {};
			\vertex (2) [above left=25pt of 1];
			\vertex (3) [below left=25pt of 1];
			\vertex (4) [above right=25pt of 1];
			\vertex (5) [below right=25pt of 1];
			\propag[glu] (1) to [edge label=$A$] (2);
			\propag[pho] (1) to [edge label=$\bar{\chi}$] (3);
			\propag[glu] (1) to [edge label=$A$] (4);
			\propag[gho] (1) to [edge label=$c$] (5);
		\end{feynhand}
	\end{tikzpicture}
\quad .
\label{eq.vertices}
\end{align}

\noindent Using these diagrams, one identifies a kind of \textit{cascade effect} in which the number of internal $A$-legs always increases when trying to construct loop Feynman diagrams, according to the diagram below,
\begin{equation}
	\begin{split}
	\begin{tikzpicture}
		\begin{feynhand}
		\vertex[dot] (1) {};
		\vertex (2)  [right=18pt of 1];
		\vertex (3) [right=12pt of 2];
		\vertex (2^) [above=2pt of 2];
		\vertex (3^) [above=2pt of 3];
		\vertex[dot] (4) [right=12pt of 2] {};
		\vertex (5) [above right=18pt of 4];
		\vertex (6) [above right=12pt of 5];
		\vertex (5^) [above left=2pt of 5];
		\vertex (6^) [above left=2pt of 6];
		\vertex[dot] (7) [above right=12pt of 5] {};
		\vertex (8) [above=18pt of 7];
		\vertex (9) [above=12pt of 8];
		\vertex (8^) [left=2pt of 8];
		\vertex (9^) [left=2pt of 9];
		\vertex[dot] (10) [above=12pt of 8] {};
		\vertex (11) [above left=18pt of 10];
		\vertex (12) [above right=18pt of 10];
		\vertex[particle] [above=18pt of 10] {$\vdots$};
		\vertex (13) [right=18pt of 7];
		\vertex (14) [right=12pt of 13];
		\vertex (13^) [above=2pt of 13];
		\vertex (14^) [above=2pt of 14];
		\vertex[dot] (15) [right=12pt of 13] {};
		\vertex (16) [above right=18pt of 15];
		\vertex (17) [below right=18pt of 15];
		\vertex[particle] [right=15pt of 15] {$\cdots$};
		\vertex (18) [below right=18pt of 4];
		\vertex (19) [below right=12pt of 18];
		\vertex (18^) [above right=2pt of 18];
		\vertex (19^) [above right=2pt of 19];
		\vertex[dot] (20) [below right=12pt of 18] {};
		\vertex (21) [right=18pt of 20];
		\vertex (22) [right=12pt of 21];
		\vertex (21^) [above=2pt of 21];
		\vertex (22^) [above=2pt of 22];
		\vertex[dot] (23) [right=12pt of 21] {};
		\vertex (24) [above right=18pt of 23];
		\vertex (25) [below right=18pt of 23];
		\vertex[particle] [right=15pt of 23] {$\cdots$};
		\vertex (26) [below=18pt of 20];
		\vertex (27) [below=12pt of 26];
		\vertex (26^) [right=2pt of 26];
		\vertex (27^) [right=2pt of 27];
		\vertex[dot] (28) [below=12pt of 26] {};
		\vertex (29) [below right=18pt of 28];
		\vertex (30) [below left=18pt of 28];
		\vertex[particle] [below=15pt of 28] {$\vdots$};
		\graph{(1) --[glu] (2) --[plain] (3)};
		\propag[plain] (2^) to (3^);
		\graph{(4) --[glu] (5) --[plain] (6)};
		\propag[plain] (5^) to (6^);
		\graph{(7) --[glu] (8) --[plain] (9)};
		\propag[plain] (8^) to (9^);
		\propag[glu] (10) to (11);
		\propag[glu] (10) to (12);
		\graph{(7) --[glu] (13) --[plain] (14)};
		\propag[plain] (13^) to (14^);
		\propag[glu] (15) to (16);
		\propag[glu] (15) to (17);
		\graph{(4) --[glu] (18) --[plain] (19)};
		\propag[plain] (18^) to (19^);
		\graph{(20) --[glu] (21) --[plain] (22)};
		\propag[plain] (21^) to (22^);
		\propag[glu] (23) to (24);
		\propag[glu] (23) to (25);
		\graph{(20) --[glu] (26) --[plain] (27)};
		\propag[plain] (26^) to (27^);
		\propag[glu] (28) to (29);
		\propag[glu] (28) to (30);
		\end{feynhand}
	\end{tikzpicture}
	\;.
	\label{CE1}
	\end{split}
\end{equation}
This makes it impossible to close loops without using the $\langle AA \rangle$ propagator\footnote{The formal proof of this result can be found in \cite{Junqueira:2018xgl}.}, which vanishes by means of \eqref{aa}. Note that, internally, the $A$-leg always propagates to the vertex $BAA$. Looking at the full action \eqref{fullaction}, the only vertex that does possess $A$-legs is $\bar{\phi}c\psi$, but the $\bar{\phi}$-leg could only propagate to the vertex $\bar{\phi}A\phi$ through $\langle\bar{\phi}\phi\rangle$; the $c$-leg only to $\bar{c}A c$ through $\langle \bar{c}c\rangle$ and; the $\psi$-leg to the vertexes $\bar{\chi}A \psi$, $\bar{\chi}cA$ or $\bar{\chi}c AA$ through $\langle \psi\bar{\chi}\rangle$ ($\langle \bar{\eta}\psi\rangle$ is not considered because there is no vertex containing $\bar{\eta}$). All possible branches produce at least one remaining internal $A$-leg, and the cascade effect is not avoided, as represented by the diagrams 
\begin{equation}
	\begin{tikzpicture}
		\begin{feynhand}
			\vertex[dot] (0) {};
			\vertex (1);
			\vertex (1^) [above=2pt of 1];
			\vertex (1<) [left=25pt of 1];
			\vertex (1<^) [above=2pt of 1<];
			\vertex[dot] (1') [left=25pt of 1] {};
			\vertex (1<^<) [left=1pt of 1<^];
			\vertex[dot] (2) [above right=25pt of 1] {};
			\vertex (4) [above=25pt of 2];
			\vertex (5) [right=25pt of 2];
			\vertex[dot] (3) [below right=25pt of 1] {};
			\vertex (6) [right=25pt of 3];
			\vertex (7) [below=25pt of 3];
			\vertex (8) [below left=25pt of 1'];
			\vertex (9) [above left=25pt of 1<^<];
			\vertex (99) [below left=2pt of 1<^<];
			\vertex (9v) [above left=25pt of 99];
			\propag[gho] (1) to (1<);
			\propag[gho] (1^) to (1<^);
			\propag[pho] (1) to (2);
			\propag[gho] (1) to (3);
			\propag[glu] (2) to (4);
			\propag[gho] (2) to (5);
			\propag[glu] (3) to (6);
			\propag[gho] (3) to (7);
			\propag[glu] (1') to (8);
			\propag[gho] (1<^<) to (9);
			\propag[gho] (99) to (9v);
		\end{feynhand}
	\end{tikzpicture}
\quad ,\quad\quad\quad 
	\begin{tikzpicture}
		\begin{feynhand}
			\vertex[dot] (0) {};
			\vertex (1);
			\vertex (1^) [above=2pt of 1];
			\vertex (1<) [left=25pt of 1];
			\vertex (1<^) [above=2pt of 1<];
			\vertex[dot] (1') [left=25pt of 1] {};
			\vertex (1<^<) [left=1pt of 1<^];
			\vertex[dot] (2) [above right=25pt of 1] {};
			\vertex (4) [above=25pt of 2];
			\vertex (5) [right=25pt of 2];
			\vertex[dot] (3) [below right=25pt of 1] {};
			\vertex (6) [right=25pt of 3];
			\vertex (7) [below=25pt of 3];
			\vertex (8) [below left=25pt of 1'];
			\vertex (9) [above left=25pt of 1<^<];
			\vertex (99) [below left=2pt of 1<^<];
			\vertex (9v) [above left=25pt of 99];
			\propag[gho] (1) to (1<);
			\propag[gho] (1^) to (1<^);
			\propag[pho] (1) to (2);
			\propag[gho] (1) to (3);
			\propag[glu] (2) to (4);
			\propag[pho] (2) to (5);
			\propag[glu] (3) to (6);
			\propag[gho] (3) to (7);
			\propag[glu] (1') to (8);
			\propag[gho] (1<^<) to (9);
			\propag[gho] (99) to (9v);
		\end{feynhand}
	\end{tikzpicture}
\quad ,\quad\quad\quad 
	\begin{tikzpicture}
		\begin{feynhand}
			\vertex[dot] (0) {};
			\vertex (1);
			\vertex (1^) [above=2pt of 1];
			\vertex (1<) [left=25pt of 1];
			\vertex (1<^) [above=2pt of 1<];
			\vertex[dot] (1') [left=25pt of 1] {};
			\vertex (1<^<) [left=1pt of 1<^];
			\vertex[dot] (2) [above right=25pt of 1] {};
			\vertex (4) [above right=25pt of 2];
			\vertex (5) [below right=25pt of 2];
			\vertex (5') [above left=25pt of 2];
			\vertex[dot] (3) [below right=25pt of 1] {};
			\vertex (6) [right=25pt of 3];
			\vertex (7) [below=25pt of 3];
			\vertex (8) [below left=25pt of 1'];
			\vertex (9) [above left=25pt of 1<^<];
			\vertex (99) [below left=2pt of 1<^<];
			\vertex (9v) [above left=25pt of 99];
			\propag[gho] (1) to (1<);
			\propag[gho] (1^) to (1<^);
			\propag[pho] (1) to (2);
			\propag[gho] (1) to (3);
			\propag[glu] (2) to (4);
			\propag[gho] (2) to (5);
			\propag[glu] (2) to (5');
			\propag[glu] (3) to (6);
			\propag[gho] (3) to (7);
			\propag[glu] (1') to (8);
			\propag[gho] (1<^<) to (9);
			\propag[gho] (99) to (9v);
		\end{feynhand}
	\end{tikzpicture}
\quad .
\end{equation}
The apparently only non-zero correlation functions are of the type
\begin{equation}
\langle BBB\ldots bb\rangle=\langle s\bar{\chi}BB \ldots bb\rangle=\langle s(\bar{\chi}BB \ldots bb)\rangle\;,\label{prop2a}
\end{equation}
\textit{i.e.}, with external $B_{\mu\nu}^a$ or $b^a$ fields. But \eqref{prop2a} automatically vanishes as it is BRST-exact. 

In a few words, using perturbative techniques, one sees that the tree-level exactness of the BS in the self-dual gauges is a consequence of the vector supersymmetry and BRST symmetry.

\subsection{Renormalization ambiguity}

Once we have at our disposal all Ward identities, we are able to construct the most general counterterm $\Sigma^c$ that can absorb the divergences arising in the evaluation of Feynman graphs. Due to the triviality of the BRST cohomology \cite{Brandhuber:1994uf,Junqueira:2017zea}, $\Sigma^c$ belongs to trivial part of the BRST cohomology. The fact that the BS theory is quantum stable is a well-known result in literature \cite{WerneckdeOliveira:1993pa, Brandhuber:1994uf,Junqueira:2017zea}. In \cite{Junqueira:2017zea}, it was introduced an extra non-linear bosonic symmetry that relates the topological ghost with the Faddeev-Popov one (among other transformations involving other fields) through the transformation
\begin{equation} \label{FPtopoghosts}
    \delta \psi^a_\mu  \longmapsto D^{ab}_\mu c^b\;,
\end{equation}
described by the Ward identity $\mathcal{T}$ in Equation \eqref{cnl1}. Taking into account this extra symmetry, from the multiplicative redefinition of the fields, sources and parameters of the model,
\begin{eqnarray} \label{Zfields}
\Phi_0 &=& Z^{1/2}_\Phi \Phi\;, \quad  \Phi_0 = \{A^a_\mu, \psi^a_\mu, c^a, \bar{c}^a, \phi^a, \bar{\phi}^a, b^a, \bar{\eta}^a, \bar{\chi}^a_{\mu\nu}, B^a_{\mu\nu}\}\;,\nonumber\\
\mathcal{J}_0 &=& Z_\mathcal{J}\mathcal{J}\;, \quad \;\;\mathcal{J} = \{\tau^a_\mu, \Omega^a_\mu, E^a, L^a, \Lambda^a_{\mu\nu}, K^a_{\mu\nu}\}\;,\nonumber\\
g_0 &=& Z_g g\;,
\end{eqnarray}
one proves the quantum stability of the BS theory in self-dual gauges with only one independent renormalization parameter, \textit{i.e.}, that the quantum action $\Gamma \equiv \Sigma(\Phi_0, \mathcal{J}_0, g_0)$ at one-loop is of the form
\begin{equation}
\Sigma(\Phi_0,\mathcal{J}_0,g_0)=\Sigma(\Phi,\mathcal{J},g)+\epsilon\Sigma^c(\Phi,\mathcal{J},g)\;,\label{abs1}
\end{equation}
with
\begin{equation}
\Sigma^c = a\int d^4x \,\left(B^a_{\mu\nu} F^a_{\mu\nu} - 2\bar{\chi}^a_{\mu\nu} D^{ab}_\mu\psi^b_\nu - g f^{abc} \bar{\chi}^a_{\mu\nu}c^bF^c_{\mu\nu}\right)\;,\label{ctfinal}
\end{equation}
whereby the resulting $Z$ factors obey the following system of equations:
\begin{eqnarray}
Z^{1/2}_A &=&Z_b^{-1/2}=Z_g^{-1}\;,\nonumber\\ Z^{1/2}_{\bar{c}}&=&Z^{1/2}_{\bar{\eta}}= Z_\psi^{-1/2} = Z_\Omega = Z^{-1/2}_c \;,\nonumber\\
Z_{\bar{\phi}}^{1/2}&=&Z_\phi^{-1/2}=Z_\tau=Z_L=Z^{-1}_gZ^{-1}_c\;,\nonumber\\ Z_E&=&Z_g^{-2}Z^{-3/2}_c\;,\nonumber\\
Z_K&=&Z_g^{-1}Z_c^{-1/2}Z_{\bar{\chi}}^{-1/2}\;,\nonumber\\
Z_\Lambda &=&Z_g^{-2}Z_c^{-1}Z_{\bar{\chi}}^{-1/2}\;,\nonumber\\
Z^{1/2}_B Z^{1/2}_A &=& Z^{1/2}_{\bar{\chi}} Z^{1/2}_c  =1 + \epsilon a\;,\label{Z1a}
\end{eqnarray}
with the independent renormalization parameter denoted by $a$. Due to the recursive nature of algebraic renormalization \cite{Piguet:1995er}, the results \eqref{Z1a} show that the model is renormalizable to all orders in perturbation theory. 

From the algebraic analysis so far, we cannot prove that $Z_g=1$, as suggested by the tree-level exactness obtained via the study of the Feynman diagrams. The system of $Z$ factors \eqref{Z1a} is undetermined. As we can easily see, the number of equations $n$ and the number of variables $z$ (the $Z$ factors) are related by $z=n+2$, indicating that there is a kind of freedom in the choice of two of the $Z$ factors. 

In \cite{Junqueira:2018zxr}, the origin of such an ambiguity was explained: it is due to the absence of a kinetic gauge field term out from the trivial BRST cohomology, and due to the absence of discrete symmetries involving the ghost fields. The symmetries of the SDL gauges eliminate the kinetic term of the Faddeev-Popov ghost in the counterterm, \textit{i.e.},
\begin{equation} \label{Zc}
Z_c Z_{\bar{c}}=1\;.
\end{equation}
Moreover, from the gauge-ghost vertex ($\bar{c}Ac$), which is also absent in the counterterm, we achieve 
\begin{equation}\label{ZgA}
Z_g Z^{1/2}_A =1\;.
\end{equation}
The two relations \eqref{Zc} and \eqref{ZgA} are decoupled, in other words, only by determining $Z_c$ or $Z_{\bar{c}}$ we do not get any information about $Z_g$ or $Z_A$. As there are no kinetic terms for the gauge field in the classical action \eqref{fullaction}, the independent determination of $Z_A$ becomes impossible. The same analysis can be performed for the bosonic and topological ghosts, see \cite{Junqueira:2018zxr}.

Extra information is then required in order to determine the system \eqref{Z1a}. In the ordinary Yang-Mills theories (quantized in the Landau gauge), $Z_c = Z_{\bar{c}}$ which relies on the discrete symmetry
\begin{eqnarray}\label{cbarc}
c^a &\longrightarrow & \bar{c}^a\;,\nonumber \\
\bar{c}^a &\longrightarrow & -{c}^a\;.
\end{eqnarray}
This condition, together with the Faddeev-Popov ghost kinetic term, are sufficient to determine $Z_c$ and $Z_{\bar{c}}$. It is easy to see that the action \eqref{fullaction} does not obey such a symmetry. Discrete symmetries between the other ghosts of topological Yang-Mills theories ($\phi^a$ and $\bar{\phi}^a$ and; $\psi^a_\mu$ and $\bar{\chi}^a_{\mu\nu}$) are also not present in \eqref{fullaction}, which explains the second ambiguity. In Witten's theory, such an ambiguity will not appear by this reasoning since Witten's action contains discrete symmetries ensured by the time-reversal symmetry \eqref{cbarc} in Landau gauge, together with 
\begin{eqnarray}\label{discrete}
    \phi &\rightarrow & \bar{\phi} \;, \quad \bar{\phi} \rightarrow \phi\;,\nonumber \\ 
    \psi_\mu &\rightarrow & \chi_\mu\;, \quad \chi_\mu \rightarrow \psi_\mu\;,
\end{eqnarray}
whereby the components of $\chi_\mu$ are defined as follows
\begin{equation}
\chi_0  \equiv \eta\;, \quad \chi_{i} \equiv \chi_{0i} = \frac{1}{2}\varepsilon_{ijk} \chi_{jk}\;,
\end{equation}
implying a ``particle-antiparticle" relationship between $\bar{c}$ and $c$, $\bar{\phi}$ and $\phi$, and $\psi_\mu$ and $\chi_\mu$, as demonstrated in \cite{Brooks:1988jm}.

This ambiguity is also present in a generalized class of renormalizable gauges \cite{Junqueira:2018zxr}. In fact, one could relate this ambiguity with the fact that all local degrees of freedom are non-physical (\textit{e.g.} the gauge field propagator is totally gauge dependent). In self-dual Landau gauges, where the vector supersymmetry is present, the Feynman diagram structure indicates that imposing $Z_c = Z_{\bar{c}}$ and $Z_\phi = Z_{\bar{\phi}}$ is consistent with the model. Hence the Z-factor system \eqref{Z1a} would naturally yield $Z_g = 1$, in accordance with the absence of radiative corrections in this gauge choice. However, without recovering the discrete symmetries between the ghosts, such an imposition seems to be artificial. As we will see later, the renormalization ambiguity can be solved in the SDL gauges, \textit{i.e.}, the discrete symmetries can be reconstructed, due to the triviality of the Gribov copies \cite{Dudal:2019bjh}, which allows for a non-local transformation with trivial Jacobian, capable of recovering such symmetries.

\section{Perturbative $\beta$ functions}\label{Beta}

Our aim in this section is to compare the DW and BS $\beta$-functions to prove that these topological gauge theories are not completely equivalent at the quantum level, and then identify  in which energy regimes the correspondence could occur. The DW $\beta$-function is well known \cite{Brooks:1988jm,Blasi:2000qw}, as we will briefly describe. It remains the task of determining the self-dual BS one to perform the comparison. 

\subsection{Twisted $N=2$ super-Yang-Mills theory}\label{BetaDW}

In \cite{Brooks:1988jm} the authors have computed the one-loop $\beta$-function of the DW theory.  Later, the authors of \cite{Blasi:2000qw} employed the algebraic renormalization techniques to also study DW theory, and prove that the $\beta$-function of twisted $N=2$ SYM ($\beta^{N=2}_g$) is one-loop exact. The reason is that the composite operator $\text{Tr}\phi^2(x)$ does not renormalize \cite{Lemes:2000ni}. For that, they considered the fact that the operator $d_{\mu\nu}$, defined in expression \eqref{Qdeltamunu}, is redundant \cite{Fucito:1997xm}. Thence, the definition of an extended BRST operator, namely, 
\begin{equation}
\mathcal{S} = s_{YM} + \omega \delta + \varepsilon_\mu \delta_\mu\;,\label{sextended}
\end{equation}
could be employed. In expression \eqref{sextended}, $\omega$ and $\varepsilon_\mu$ are global ghosts, and $\delta$ and $\delta_\mu$ were defined in equations \eqref{Qdelta} and \eqref{Qdeltamu}. The relevant property of the operator $\mathcal{S}$ is that it is on-shell nilpotent in the space of integrated local functionals, since
\begin{equation} \label{extendedBRST}
\mathcal{S}^2 =  \omega\varepsilon_\mu\partial_\mu + \text{eqs of motion}\;.   
\end{equation}
We point out that this extended BRST construction requires the equations of motion to obtain a nilpotent BRST operator---a standard behavior of Witten theory, representing a different quantization scheme of the BS theory. Considering the non-renormalization of $\text{Tr}\phi^2$ and the on-shell cohomology of the operator defined in eq. \eqref{extendedBRST}, the result is that the $\beta$-function only receives contributions to one-loop order, and is given by
\begin{equation} \label{betafunctionW}
\beta_g^{N=2} = - Kg^3\;,
\end{equation}
with $K$ being a constant. The computation of $\beta_g^{N=2}$ via Feynman diagrams was performed in \cite{Brooks:1988jm} by evaluating the one-loop contributions to the gauge field propagator (where the Landau gauge was used to fix the Yang-Mills  symmetry of Witten action \eqref{gaugesymmetryWitten}). The behavior of one-loop exactness of the $N=2$ $\beta$-function had been usually understood in terms of the analogous Adler-Bardeen theorem for the $U(1)$ axial current in the $N=2$ SYM \cite{West:1990}. 

Despite the independence of the Witten partition function under changes in the coupling constant, the result \eqref{betafunctionW} should not be surprising. In the twisted version, we can see that the trace of the energy-momentum is not zero, but given by (see \cite{Witten:1988ze})
\begin{equation}
g_{\mu\nu}T^{\mu\nu} = \text{Tr}\{D_\mu \phi D^\mu \bar{\phi} -2i D_\mu\eta \psi^\mu + 2i \bar{\phi}[\psi_\mu, \psi^\mu] +2i \phi[\eta, \eta] +\frac{1}{2}[\phi, \bar{\phi}]^2] \}\;,
\end{equation}
meaning that $S_W$ is not conformally invariant under the transformation
\begin{equation}
\delta g_{\mu\nu} = h(x) g_{\mu\nu}\;,
\end{equation}
for an arbitrary real function $h(x)$ on $M$. Nonetheless, the trace of the energy-momentum tensor can be written as a total covariant divergence,
\begin{equation}
g_{\mu\nu}T^{\mu\nu} = D_\mu\left[ \text{Tr}( \bar{\phi}D^\mu \phi -2i \eta \psi^\mu)\right]\;,
\end{equation}
which means that $S_W$ is invariant under a global rescaling of the metric, $\delta g_{\mu\nu} = w g_{\mu\nu}$, with $w$ constant  \cite{Witten:1988ze}. The $N=2$ $\beta$-function only vanishes if we take the weak coupling limit $g^2 \rightarrow 0$, 
\begin{equation}
\beta_g^{N=2}(g^2 \rightarrow 0) = 0\;.
\end{equation}
In this limit, the possibility of loop corrections to the effective action is eliminated, and the Donaldson polynomials are reproduced as the observables of the theory. There is no Ward identity, or a particular property of the vertices and propagators of $S_W$, capable of eliminating these quantum corrections for an arbitrary energy regime---this situation is distinct from the BS theory in the self-dual Landau gauges.  

\subsection{Baulieu-Singer topological theory}\label{BetaBS}

As suggested by the tree-level exactness of the BS theory in the self-dual Landau gauges, according to the analysis of the Feynman diagrams performed in Section \ref{SDLgauges}, we will formally prove that the self-dual BS theory is conformal. Before proving the vanishing of the BS $\beta$-function in this gauge, we will first discuss the non-physical character of the coupling constant in this off-shell approach, since $g$ is introduced in the BS theory as a gauge parameter, in the trivial part of the BRST cohomology. 

\subsubsection{Nonphysical character of the $\beta$ function in the off-shell approach}\label{Nonphysical}
 
In \cite{Brooks:1988jm}, Brooks \emph{et al.} argued that only one counterterm is required in the on-shell Witten theory, specifically for the YM term $\text{Tr}\,F^2_{\mu\nu}$. In any case, the Donaldson invariants are described by DW theory in the weak coupling limit $g^2 \rightarrow 0$, where the theory is dominated by the classical minima.  On the other hand, it is evident that the BS theory is distinct from Brooks \textit{et al.} construction because their methods are based on different BRST quantization schemes, with different cohomological properties. We do not expect a similar result in the BS theory. According to the cohomology of the BS model, if the $\beta^{BS}_g$ is not zero, we should find that it is $\text{Tr}\,(F_{\mu\nu} \pm \widetilde{F}_{\mu\nu})^2$ rather than $\text{Tr}\, F^2_{\mu\nu}$ which is renormalized\footnote{See \cite{Birmingham:1988ap}, where Birmingham \textit{et al.} had employed the Batalin-Vilkovisky algorithm \cite{Batalin:1984jr}---a similar quantization to BS approach, \textit{i.e.}, with similar cohomological properties.}. In this way, the minima of the effective action preserves the instanton configuration at the quantum level, according to the global degrees of freedom of the instantons, which defines the observables of the BS theory---the Donaldson invariants. 

A possible discrepancy between $\beta$-functions for the BS approach in different gauge choices cannot be attributed to a gauge anomaly, since it is forbidden in these models due to the trivial BRST cohomology \cite{Birmingham:1988ap}, cf. equation \eqref{anomalycondition}. For instance, if we would had chosen the gauge $D^{ab}_\mu\psi^b_\mu = 0$ for the topological ghost, with the covariant derivative instead of the ordinary one, the vector supersymmetry would be broken, and the gauge propagator would not vanish to all orders anymore. In ordinary Yang-Mills theories, the $\beta$-function is an on-shell gauge-invariant physical quantity. Nonetheless, in gauge-fixed BRST topological theories of BS type, \textit{the coupling constant is non-physical}, introduced in the trivial part of  the cohomology, together with the gauge-fixing action. In these terms, it is not contradictory that the $\beta$-function is gauge dependent as it computes the running of a non-physical parameter. We must observe that the physical observables of the theory, the Donaldson invariants, naturally do not depend on the gauge coupling.  So that there is no inconsistency that the observables of this kind of theory, described by topological invariants, \textit{i.e.}, exact numbers, do not depend on the coupling constant, and consequently on its running, being $g$ an unobservable quantity.

As DW and BS theories possess the same observables, we should then consider the instanton configuration not as a gauge fixing condition, but as a physical requirement in order to obtain the correct degrees of freedom that correspond to the description of all global observables. Furthermore, the Atiyah-Singer index theorem \cite{Atiyah:1963zz} determines the dimension of the instanton moduli space, in which the Donaldson invariants are defined---see \cite{Witten:1976ck,Jackiw:1976fs} for some exact instanton solutions, whose properties cannot be attributed to gauge artifacts.

\subsubsection{Conformal structure of the self-dual gauges}\label{ConformalSDL}

To prove that the algebraic renormalization is in harmony with the Feynman diagram analysis in the self-dual Landau gauges, which shows that the BS model does not receive radiative corrections in this gauge, we must invoke a result recently published in \cite{Dudal:2019bjh}. In this work, it was demonstrated that the Gribov ambiguities \cite{Gribov:1977wm,Sobreiro:2005ec} are inoffensive in the self-dual BS theory\footnote{The result was proved to be valid to all orders in perturbation theory by making use of the Zwanziger's approach \cite{Zwanziger:1988jt} to the Gribov problem \cite{Gribov:1977wm}.}. The quantization of this model in a local section of the field space where the eigenvalues of the Faddeev-Popov determinant are positive, is equivalent to its quantization in the whole field space. In other words, the introduction of the Gribov horizon does not affect the dynamics of the BS theory in SDL gauges, as its correspondent gap equation forbids the introduction of a Gribov massive parameter in the gauge field propagator. This result also suggests that the fiber bundle structure of the BS theory is trivial \cite{Singer:1978dk}.

Let us quickly recall the Gribov procedure in the quantization of non-Abelian gauge theories \cite{Gribov:1977wm,Sobreiro:2005ec}. It essentially consists in eliminating remaining gauge ambiguities usually present in non-Abelian gauge theories, known as Gribov copies, which are not eliminated in the Faddeev-Popov (FP) procedure \cite{Faddeev:1967fc,Itzykson:1980rh}. In Yang-Mills theories, the FP gauge-fixing procedure results in the well-known functional generator
\begin{equation} \label{ZwithFPdet}
Z_{YM} = \mathcal{N} \int \mathcal{D}A \vert \det[ -\partial_\mu D_\mu^{ab}] \vert \delta (\partial_\mu A_\mu) e^{-S_{YM}}= \mathcal{N} \int \mathcal{D}A\mathcal{D}\bar{c}\mathcal{D}c e^{-(S_{YM}+S_{gf})}\;,
\end{equation}
whereby $S_{gf}$ is the well-known gauge-fixing action given by
\begin{equation} \label{SgfYM}
    S_{gf} = \int d^4x \left(\bar{c}^a \partial_\mu D_\mu^{ab} c^b - \frac{1}{2\alpha} (\partial_\mu A_\mu^a)^2\right)\;. 
\end{equation}
In \eqref{SgfYM}, the limit $\alpha\longrightarrow0$ must be taken in order to reach the Landau gauge,
\begin{equation}
    \partial_\mu A_\mu=0\;.\label{landau}
\end{equation}
Consider a gauge orbit\footnote{The gauge orbit is the equivalence class of gauge field configurations that only differ by a gauge transformation, representing thus the same physics according to the gauge the gauge principle.}
\begin{equation} \label{A^U}
A_\mu^U = U A_\mu U^{\dagger} - \frac{i}{g} (\partial_\mu U) U^{\dagger}\;, 
\end{equation}
with $U = e^{-ig T^a \theta^a(x)}\;\Big|\;U \in SU(N)$ with $\theta^a(x)$ being the local gauge parameters of the non-Abelian symmetry, and $T^a$ the generators of the gauge group. The FP hypothesis \cite{Faddeev:1967fc,Itzykson:1980rh} is that there is only one gauge configuration in the orbit \eqref{A^U} obeying the Landau gauge condition \eqref{landau}. In his seminal work \cite{Gribov:1977wm}, V.~N.~Gribov demonstrated that this hypothesis fail at the YM low energy regime because one can always find two configurations $\tilde{A}$ and $A$ obeying the Landau gauge condition and yet being related by a gauge transformation. At infinitesimal level, the condition for a configuration $A$ to have a Gribov copy $\tilde{A}$ is that the FP operator develops zero-modes through
\begin{equation} \label{copyeqYMGribov}
-\partial_\mu D_\mu \theta = 0\,.
\end{equation}
with $\theta^a$ taken as an infinitesimal parameter, $U \approx 1 - \theta^a T^a$. Equation \eqref{copyeqYMGribov} is recognized as the Gribov copies equation in the Landau gauge (and also in linear covariant gauges--See \cite{Sobreiro:2005vn,Capri:2015pja,Capri:2015ixa,Capri:2015nzw}). Equation \eqref{copyeqYMGribov} can be seen as an eigenvalue equation for the FP operator where $\theta$ is the zero mode of the operator. In Landau gauge, this operator is Hermitian, and thus, its eigenvalues are real. For values of $A_\mu$ sufficiently small, the eigenvalues of the FP operator will be positive, as $-\partial^2$ only has positive eigenvalues\footnote{In Abelian theories, such as QED, $-\partial^2$ is the ``FP operator", and the copy equation only possesses trivial solutions in the thermodynamic limit, meaning that the Gribov copies are inoffensive in this case.}. As $A_\mu$ increases, it will attain a first zero mode \eqref{copyeqYMGribov}. Such region in which the FP operator has its first vanishing eigenvalue is called Gribov horizon (See also \cite{Sobreiro:2005ec}). Gribov's proposal was to restrict the path integral domain to the region $\Omega$ (the Gribov region) defined by
\begin{equation}
\label{GribovOmega}
\Omega = \{A^a_\mu; \; \partial_\mu A_\mu = 0, \; -\partial D > 0 \}\,.
\end{equation}
Such restriction ensures the elimination of all infinitesimal copies and also guarantees that no independent field is eliminated \cite{DellAntonio:1991mms}.

The implementation of the restriction to the Gribov region $\Omega$ is accomplished by the introduction of a step-function $\Theta(-\partial D)$ in the Feynman path integral, that leads to the well-known \textit{no-pole condition} for the FP ghost propagator $\langle(\partial D)^{-1}\rangle$, which explodes at when a zero mode is attained. The main result of introducing the restriction of the Feynman path integral domain to the Gribov region is a modified gluon propagator, due to the emergence of a massive parameter for the gauge field, the so called Gribov parameter $\gamma$. In the presence of the Gribov horizon, the gluon propagator takes the form
\begin{equation} \label{gluonpropGribov}
    \langle A^a_\mu(k) A_\nu^b(k) \rangle = \delta^{ab} \delta(p+k) \frac{k^2}{k^4 + \gamma^4} P_{\mu\nu}(k)\;,
\end{equation}
where $P_{\mu\nu}(k) = \delta_{\mu\nu} - k_\mu k_\nu/k^2$, and $\gamma$ fixed by the gap equation \cite{Zwanziger:1988jt, Vandersickel:2010ti}, 
\begin{equation}
\frac{\partial \Gamma}{\partial \gamma^2} = 0\;.
\end{equation}

According to Zwanziger's generalization \cite{Zwanziger:1988jt}, the gap equation above is valid to all orders in perturbation theory---see \cite{Gomez:2009tj, Capri:2012wx}, where the all-order proof of the equivalence between Gribov and Zwanziger methods was worked out. An important feature of the Gribov parameter is that it only affects the infrared dynamics. The matching between Gribov-Zwanziger theory and recent lattice data is achieved through the introduction of two-dimensional condensates, see \cite{Dudal:2008sp}.  The introduction of the Gribov horizon in the action explicitly breaks the BRST symmetry. This is usually an unwanted result, as the BRST symmetry is necessary to prove the unitarity, to ensure the renormalizability to all orders, and to define the physical gauge-invariant observables of the theory \cite{Slavnov:1989jh,
Frolov:1989az, Dudal:2009xh}. This breaking however brought to light the physical meaning of the infrared $\gamma$ parameter, and its intrinsic non-perturbative character. One can prove that the BRST breaking is proportional to $\gamma^2$, in other words, the BRST symmetry is restored in the perturbative regime. One says that the BRST symmetry is only broken in a \textit{soft} way, cf. \cite{Baulieu:2008fy, Sorella:2009vt, Dudal:2009xh, Sorella:2011tu}. Only more recently, a universal, gauge independent, (non-perturbative) BRST invariant way to introduce the Gribov horizon was developed \cite{Capri:2015ixa,Capri:2015nzw,Capri:2016aqq,Pereira:2016fpn,Capri:2018ijg}.

In the self-dual topological BS theory, it was proved in \cite{Dudal:2019bjh} that all topological gauge copies associated to the gauge ambiguities \eqref{eqn:gluon-symmetry} and \eqref{eqn:top-parameter-symmetry}, are eliminated through the introduction of the usual Gribov restriction, in which the path integral domain is restricted to the region $\Omega$---see eq. \eqref{GribovOmega}. Moreover, due to the triviality of the gap equation, it was verified that the Gribov copies does not affect the infrared dynamics of the self-dual BS theory because $\gamma_{BS} = 0$ is the only possible solution of the gap equation\footnote{A similar situation occurs in the $N=4$ SYM which possesses a vanishing $\beta$-function, indicating the conformal structure of the self-dual BS. The absence of an invariant scale makes it impossible to attach a dynamical meaning to the Gribov parameter \cite{Capri:2014tta}.}. Thus, no mass parameter seems to emerge in the BS theory, preserving its conformal character at quantum level. Specifically, the tree-level exactness in SDL gauges is preserved. Such a behavior can be inferred from the absence of radiative corrections which ensures the semi-positivity of all two-point functions. The FP ghost propagator, for instance, reads
\begin{equation}
\langle \bar{c}^a(k) c^b(k) \rangle = \delta^{ab} \frac{1}{k^2}\;,
\end{equation}
which is valid to all orders, demonstrating that the FP operator will remain positive-definite at quantum level, consistent with the inverse of the FP propagator being positive, thus proving that we are inside the Gribov region. Moreover, the gauge two-point function remains trivial, \textit{i.e.}, $\langle A^a_\mu(k) A_\nu^b(k) \rangle =0$ to all orders. 

Exploring the positive-definiteness of the FP ghost propagator, we are able to safely perform the following shifts:
\begin{eqnarray}\label{shifts}
\bar{\eta}^a&\longmapsto&\bar{\eta}^a+\bar{c}^a\;,\nonumber\\
\phi^b&\longmapsto&\phi^b-gf^{cde}(\partial_\nu D_\nu^{bc})^{-1}\partial_\mu\left(c^d\psi^e_\mu\right)\;,\nonumber\\
\bar{c}^a&\longmapsto&\bar{c}^a-\frac{1}{2}gf^{cde}\bar{\chi}^d_{\mu\nu}(F_\pm)^e_{\mu\nu}(\partial_\nu D_\nu^{ca})^{-1}\;.
\end{eqnarray}
It is worth noting that these shifts generate a trivial Jacobian. Calling $\frac{1}{2}\rho_1 = \alpha$ and $\frac{1}{2}\rho_2 =\beta$, and implementing the BS gauge constraints \eqref{gf1BS} and \eqref{gf3BS}, together with $\partial_\mu \psi_\mu = 0$, the final gauge-fixing action, integrating out the auxiliary fields $b$ and $B$ in the action \eqref{SgfBS}, is
\begin{eqnarray} \label{Zfp1}
S_{gf}(\alpha, \beta)&=&\int d^4x\left[-\frac{1}{2\alpha}(\partial A)^2-\frac{1}{4\beta}F_\pm^2\right]-\int d^4x\left[
\left(\bar{\eta}^a-\bar{c}^a\right)\partial_\mu\psi^a_\mu \right.\nonumber\\
&+&\left.\bar{c}^a\partial_\mu D_\mu^{ab}c^b-\frac{1}{2}gf^{abc}\bar{\chi}^a_{\mu\nu}c^b\left(F_{\mu\nu}^c\pm\widetilde{F}_{\mu\nu}^c\right)-\bar{\chi}^a_{\mu\nu}\left(\delta_{\mu\alpha}\delta_{\nu\beta}\pm\frac{1}{2}\epsilon_{\mu\nu\alpha\beta}\right)D_\alpha^{ab}\psi_\beta^b\right.\nonumber\\
&+&\left.\bar{\phi}^a\partial_\mu D_\mu^{ab}\phi^b + gf^{abc}\bar{\phi}^a\partial_\mu\left(c^b\psi^c_\mu\right)\right]\;,
\end{eqnarray}
where $F_\pm=F\pm\tilde{F}$ and $D_\pm\equiv \left(\delta_{\mu\alpha}\delta_{\nu\beta}-\delta_{\nu\alpha}\delta_{\mu\beta}\pm\epsilon_{\mu\nu\alpha\beta}\right)D_\alpha^{ab}$. The self dual Landau gauges is recovered by setting $\alpha, \beta \rightarrow 0$. Then, applying the shifts \eqref{shifts} on the action $S_{gf}(\alpha, \beta)$, one gets
\begin{eqnarray}
S_{gf}^{shifted}(\alpha,\beta)&=&\int d^4x\left[-\frac{1}{2\alpha}(\partial A)^2-\frac{1}{4\beta}F_\pm^2\right]-\int d^4x\left[
\bar{\eta}^a\partial_\mu\psi^a_\mu+\bar{c}^a\partial_\mu D_\mu^{ab}c^b\right.\nonumber\\
&-&\left.\bar{\chi}^a_{\mu\nu}\left(\delta_{\mu\alpha}\delta_{\nu\beta}\pm\frac{1}{2}\epsilon_{\mu\nu\alpha\beta}\right)D_\alpha^{ab}\psi_\beta^b+\bar{\phi}^a\partial_\mu D_\mu^{ab}\phi^b\right]\;.\label{Zfp2}
\end{eqnarray}
As the Jacobian of the shifts that performs $S_{gf}(\alpha,\beta) \rightarrow S_{gf}^{shifted}(\alpha,\beta)$ is trivial, the quantization of both actions are perturbatively equivalent, cf. \cite{Blasi:2000qw}. Such a Jacobian only generates a number that can be absorbed by the normalization factor. This shows that the discrete symmetries \eqref{cbarc} and \eqref{discrete} present in the Witten theory can be recovered, which naturally impose the relations 
\begin{equation}\label{Zrelation}
Z_c = Z_{\bar{c}} \quad \text{and} \quad Z_\phi = Z_{\bar{\phi}}
\end{equation}
to be valid in the BS theory. Hence, combining \eqref{Zrelation} with the $Z$-factor system \eqref{Z1a}, one obtains
\begin{equation}\label{ZgBS}
Z_g = 1\;,    
\end{equation}
which proves that the algebraic analysis is in harmony with the result obtained via the study of the Feynman diagrams in the presence of the vector symmetry, \textit{i.e.}, that the topological BS theory (following the self-dual Landau gauges) is conformal, in accordance with the absence of radiative corrections. 

The algebraic renormalization and the Feynman diagram analysis consist of two independent methods. In the loop expansion, used to construct the diagrams in Section \ref{Absence}, we expand around the trivial $A=0$ sector, \textit{i.e.}, around an instanton with winding number zero. One may wonder if it is physically relevant, thinking about the importance of instanton configurations in the topological theory, in order to construct the Donaldson invariants. Exactly the topological nature of the off-shell BS theory saves the day here. Let us first remark that it is possible to write down a BRST invariant version of the Gribov restriction, that is, if $\gamma$ were to be nonzero, whilst preserving equivalence with the above formalism\footnote{In the sense that all correlation functions will be identical.}, see \cite{Capri:2015ixa,Capri:2016aqq,Capri:2018ijg} for details. As already reminded before, the topological partition function does not depend on the coupling $g$. This means all observables can be computed in the $g\to0$ limit. Expanding around a nontrivial instanton background rather than around $A=0$  would lead to corrections of the type $e^{-1/g^2}$ into the effective action, but the latter vanishes exponentially fast once $g\to0$ is considered, that only represent a liberty of the theory, \text{i.e.}, it would not affect the global observables, see \cite{Dudal:2019bjh}. As such, we can a priori work around $A=0$, without losing the generality of the result, which will be unaltered for a generic instanton background. 

\section{Characterization of the DW/BS correspondence}\label{DW/BS}

We will characterize in this section the quantum correspondence between the twisted $N=2$ SYM in the ultraviolet regime and the conformal Baulieu-Singer theory in the SDL gauges. 

\subsection{Quantum equivalence between DW and self-dual BS theories}

The result obtained in \eqref{ZgBS} in the SDL gauges proves that the self-dual BS $\beta$-function vanishes. This result is completely different from the twisted $N=2$ SYM which receives one-loop corrections, and possesses a non-vanishing $\beta$-function given by \eqref{betafunctionW}. The correspondence between the BS and $N=2$ $\beta$-functions occurs when we take the weak coupling limit ($g^2 \rightarrow0)$ on the $N=2$ side. In this limit, $\beta^{N=2}_g \rightarrow 0$. On the BS side, however, the vanishing of the $\beta$-function is valid for an arbitrary coupling constant, and not only for a weak coupling, being the conformal property a consequence of the vector supersymmetry which forbids radiative corrections. In DW theory, such a property is obtained by taking $g^2 \rightarrow 0$ as small as we want (as long as $g^2 \neq 0$), as a consequence of the property that shows that the observables of DW theory are insensitive under changes of $g$. That is how Witten computed its partition function that naturally reproduces the Donaldson invariants for four manifolds, \textit{i.e.}, by eliminating the influence of the vertices at $g^2 \rightarrow 0$, and taking only the quadratic part of the action. The BS theory is also invariant under changes of $g$, as it only appears in the trivial part of the BRST cohomology, but the tree-level exactness is a general property of the BS theory in self-dual Landau gauges, \textit{i.e.}, it is valid for an arbitrary perturbative regime.  

The characterization of the correspondence between the twisted $N=2$ SYM and a conformal field theory is now complete. The fact that the twisted $N=2$ SYM in the weak coupling limit, and the BS theory share the same global observables is a well-known result in literature \cite{Delduc:1996yh, Boldo:2003ci, Boldo:2003jq}. In the DW theory, the Donaldson invariants are defined by the $\delta$-supersymmetry \eqref{Wittensym} according to the bi-descent equations encoded in \eqref{Donaldsoninv}. In the BS one, the same bi-descent equations appears, constructed from the $n$'th Chern class $\widetilde{\mathcal{W}}_n$ defined in terms of the universal curvature in the extended space $M \times \mathcal{A}/\mathcal{G}$. Such an equivalence is ensured by the equivariant cohomology that allows for the replacement $s \rightarrow \delta$, as $\widetilde{\mathcal{W}}_n$ is invariant under ordinary Yang-Mills transformations. We are now defining in which energy regimes such an equivalence occurs when we employ the self-dual Landau in the BS, and formal proving the correspondence between the twisted $N=2$ and a conformal gauge theory. The fact that the observables are the same, as a consequence of the equivariant cohomology, do not characterizes the correspondence at quantum level (we will provide a counter-example in the next section). The correspondence between the DW and BS observables, given by the equivalence  
\begin{equation}
\langle  \mathcal{O}^{DW}_{\alpha_1}(\phi_i) \mathcal{O}^{DW}_{\alpha_2}(\phi_i)  \cdots \mathcal{O}^{DW}_{\alpha_p}(\phi_i)\rangle_{g^2 \rightarrow 0} = \langle  \mathcal{O}^{BS}_{\alpha_1}(\phi^\prime_i) \mathcal{O}^{BS}_{\alpha_2}(\phi^\prime_i)  \cdots \mathcal{O}^{BS}_{\alpha_p}(\phi^\prime_i)\rangle_{SDL}\;,
\end{equation}
is independent of the gauge choice. The field content that defines the observables are the same in both theories, $\phi_i \equiv \phi^\prime_i$, since the observables are independent of the FP ghosts (which appear in the gauge-fixed BS action). In a few words, $\mathcal{O}^{DW}_{\alpha}(\phi_i) \equiv \mathcal{O}^{BS}_{\alpha}(\phi^\prime_i)$, represented by the product in eq. \eqref{Donaldsoninv}. As demonstrated in Section \ref{observables}, the BS observables naturally do not depend on $(c,\bar{c})$, due to the invariance of $\widetilde{\mathcal{W}}_n$ under $s_{YM}$ (the Yang-Mills BRST operator). The BS reproduces the Donaldson polynomials only as a consequence of the structure of the off-shell BRST transformations \eqref{brst1}. Witten works exclusively in the moduli space $\mathcal{A}/\mathcal{G}$, \textit{i.e.}, without fixing the gauge, being its observables naturally independent of the FP ghost. 

Another crucial point is that the gauge fixing term in which the FP ghosts are introduced in the self-dual BS theory does not allow for the influence of Gribov copies. For this reason, working in the moduli space $\mathcal{A}/\mathcal{G}$ in the DW theory are completely correspondent to work in the BS theory in SDL gauges for an arbitrary $g$, since $\gamma^4 = 0$. Fixing the remaining YM gauge symmetry of Witten's action \eqref{gaugesymmetryWitten}, instead of working in $\mathcal{A}/\mathcal{G}$, would not break such a correspondence since the Gribov copies could only affect the non-perturbative regime, being inoffensive at the ultraviolet limit $g^2 \rightarrow 0$.  The quantum equivalence are illustrated by the agreement between the $\beta$- functions, $\beta^{N=2}_g(g^2 \rightarrow 0) = \beta_g^{BS}(g) = 0$.

Finally, due to the property of Witten theory \eqref{Wproperty}, which ensures that Witten theory can be extended to any Riemannian manifold, the DW/BS correspondence is characterized as follows: the twsited $N=2$ SYM at $g^2 \rightarrow 0$, in any Riemannian manifold (that can be continuously deformed into each other, including $\mathbb{R}^4$)\footnote{This is the only requirement that guarantees that the observables of both sides are correspondent, as the conformal BS is defined in Euclidean spaces. In the DW theory, spaces that can be continuously deformed, one into the other, represent the same Donaldson invariants for a class of manifolds, since a continuous variation of the metric is equivalent to add a $\delta$-exact term to the action, which does not alter the observables.  }, defined in the instanton moduli space $\mathcal{A}/\mathcal{G}$, is correspondent to the topological BS theory in the self-dual Landau gauges in Euclidean spaces, in an arbitrary perturbative regime. Such a BS theory consists of a conformal field theory, where the gauge copies are inoffensive in the infrared, since the massive infrared parameter originated from the gauge copies vanishes in this gauge---see Table \ref{tableDW/BS} bellow.

\begin{table}[h]
\centering
   \begin{tabular}{l|c}
\textbf{Twisted $N=2$ SYM}&\textbf{BS in self-dual Landau gauges}\\\hline
On-shell $\delta$-supersymmetry& Off-shell BRST + vector supersymmetry \\
Donaldson invariants ($\delta$) & Donaldson invariants ($s\rightarrow \delta$) \\
$g^2 \rightarrow 0$&arbitrary $g$\\
Any Riemannian manifold, $g_{\mu\nu}$&Euclidean spaces, $\delta_{\mu\nu}$\\ 
$\mathcal{A}/\mathcal{G}$& gauge-fixed $\vert$ $\gamma^4_{Gribov} = 0$\\ 
$\beta_g^{N=2} \rightarrow 0$ &$\beta_g^{BS}(g) = 0$
\end{tabular}
\caption{Characterization of the DW/BS correspondence.}
\label{tableDW/BS}
\end{table}

We emphasize that we use perturbative techniques to prove the conformal property of the self-dual BS theory. The fact that the self-dual BS theory in the strong limit $g^2 \rightarrow \infty$ is also correspondent to Witten's TQFT defined at $g^2 \rightarrow 0$, can be conjectured by means of the cohomological structure of the off-shell BRST symmetry. Changing $g$ in the BS theory is equivalent to add a BRST-exact term in the action, \textit{i.e.}, it is equivalent to perform a change in the gauge choice. Moreover, the global observables of BS theory, described by the Chern classes $\widetilde{\mathcal{W}}_n$, does not depend on the gauge choice, having only the power of reproducing the Donaldson invariants for four-manifolds. Also, the Gribov ambiguities are irrelevant to the BS model (at least in the self-dual Landau gauges), a property that should remain valid at the strong regime.

\subsection{Considerations about the gauge dependence and possible generalizations}

Due to the exact nature of the topological Donaldson invariants, which are given by exact numbers, we can consider the supposition that quantum corrections should not affect the tree-level results, and that the description of the Donaldson invariants by the gauge-fixed BS approach should not depend on the gauge choice. Although intuitive, this argument is not sufficient or complete. As a counterexample, we invoke the $\beta$-function obtained by  Birmingham \textit{et al.} in \cite{Birmingham:1988ap}, where the Batalin-Vilkovisky (BV) algorithm \cite{Batalin:1984jr} was employed. Such a model possesses similar cohomological properties to the BS theory. For a particular configuration of auxiliary fields used in \cite{Birmingham:1988ap}, the BV method recovers the BS gauges with $\rho_1 = \rho_2 = 0$, together with the constraint $D_\mu \psi_\mu = 0$---see eq. \eqref{eqn:gauge-fixingsBS}. This constraint on the topological ghost, with the covariant derivative instead of the ordinary one, breaks the vector supersymmetry, allowing for quantum corrections. Consequently, the $\beta$-function computed by Birmingham \textit{et al.} is not zero. As noted by the authors of \cite{Birmingham:1988ap}, it is $\text{Tr}\,(F_{\mu\nu} \pm \widetilde{F}_{\mu\nu})^2$ rather than $\text{Tr}\, F_{\mu\nu}^2$ which is renormalized, meaning that the vacuum configurations are preserved. As expected, the structure of the instanton moduli space, that defines the Donaldson invariants, is not altered. 

About the gauge dependence of the $\beta$-function in off-shell topological gauge models, see Sec. \ref{Nonphysical}. The coupling constant in this model is non-physical, belonging to the trivial part of cohomology. Any change in the unobservable coupling constant only leads to a BRST-exact variation. The only observables are the global ones, and we must expect that, for different gauge choices, the global observables are not affected. According to the result of Birmingham \textit{et al.} in \cite{Birmingham:1988ap}, it is possible to obtain non-trivial quantum corrections without destroying the topological properties of the underlying theory, preserving the observables. Analogously, we can consider the possibility in which the fields could also be consistently renormalized, \textit{i.e.}, in such a way that the bi-descent equations, that defines the Donaldson invariants, are not altered. This reasoning shows that the renormalization of topological gauge theories, consistent with the global observables, is not a trivial issue. 

 The vector supersymmetry, that leads to the conformal BS theory, is a particular symmetry of the self-dual  Landau gauge. One must note that $\partial_\mu A_\mu = D_\mu A_\mu$, due to antisymmetric property of $f^{abc}$. To impose $\partial_\mu \psi_\mu = 0$ or $D_\mu \psi_\mu = 0$ automatically preserves the instanton moduli space, where $A_\mu$ and $\psi_\mu$ obey the same equations of motion, according to the Atiyah-Singer theorem that correctly defines its dimension. The preservation of the instanton moduli space is then the only requirement of the topological theory, being the conformal property a particular feature of the self-dual Landau gauges. The dimension of the instanton moduli space should not depend on the gauge choice, being protected by the Atiyah-Singer theorem. 
 
 The second generalization that can be worked out is in the direction of developing a model in which the BS theory can be constructed for a generic $g_{\mu\nu}$. Again, any change on the Euclidean metric to a generic one corresponds to the addition of a BRST-exact term in the BS theory. This means that such a variation can be interpreted as a change in the gauge choice, and the previous discussion can be also applied here. The vector supersymmetry is easily defined in flat spaces. In order to reproduce the conformal properties of the SDL gauges in Euclidean space for a generic $g_{\mu\nu}$, we will face the task of finding a class of metrics whose corresponding action possesses a rich set of Ward identities ($\mathbf{W}_I$), capable of reproducing the same effect of the self-dual ones, see Appendix \ref{Ap1}, given by the 11 functional operators $\mathbf{W}^{BS}_I \equiv \{\mathcal{S}, \mathcal{W}_\mu, \mathcal{W}^a_1,\mathcal{W}^a_2,\mathcal{W}^a_3,  \mathcal{W}^a_4, \mathcal{G}^a_\phi, \mathcal{G}^a_1, \mathcal{G}^a_2, \mathcal{T}, \mathcal{G}_3\}$.

Besides that, we will face another inconvenient task of having to study the Gribov copies in curved spacetimes, which is a highly nontrivial problem. The vanishing of the Gribov parameter in the self-dual BS in Euclidean spaces ensures that the DW/BS correspondence is valid for a generic coupling constant on the BS side.

\section{Conclusions}\label{Conclusions}

We perform a comparative study between Donaldson-Witten TQFT \cite{Witten:1988ze} and the Baulieu-Singer approach \cite{Baulieu:1988xs}. While DW theory is obtained via the twist transformation of the $N=2$ SYM, BS theory is based on the BRST gauge fixing of an action composed of topological invariants---see Sections \ref{DWtheory} and \ref{BStheory}, respectively. Besides that, Witten theory is defined by an on-shell supersymmetry, according to the fermionic symmetry, see eq. \eqref{Wittensym}, whose associated charge is only nilpotent if we use the equations of motion. Such a symmetry defines the observables of the theory, given by the Donaldson polynomials. The BS approach, in turn, consists of an off-shell BRST construction, which enjoys the topological BRST symmetry \eqref{brst1}, whose observables are also given by the Donaldson invariants, due to the equivariant cohomology---defined by Witten's fermionic symmetry---which also characterizes the BS observables that can be written as Chern classes for the curvature in the extended space $M \times \mathcal{A}/\mathcal{G}$, where $M$ is a Riemannian manifold and $\mathcal{A}/\mathcal{G}$ is the instanton moduli space, see Section \ref{observables}. Despite sharing the same observables, we note that these theories do not necessarily have the same quantum properties, as Witten and BS actions do not differ by a BRST-exact term, cf. \eqref{BSWittenrelation}. In a few words, the BRST quantization schemes of Witten and BS theories are not equivalent.  

In harmony with the quantum properties of BS approach in the self-dual Landau gauges, see Section \ref{SDLgauges}, we formally prove that the topological self-dual BS theory is conformal. According to the Feynman diagram analysis performed in \cite{Junqueira:2018xgl}, we proved the absence of quantum corrections in the BS theory in the presence of the vector supersymmetry. In Section \ref{Nonphysical}, we discussed the nonphysical character of the coupling constant in the off-shell BS approach. Then, to construct an algebraic proof that the self-dual BS is conformal, we first solved the renormalization ambiguity in topological Yang-Mills theories described in \cite{Junqueira:2018zxr}, using a nonlocal transformation which recovers discrete symmetries between ghost and antighost fields. Such nonlocal transformations showed to be a freedom of the self-dual BS theory due to the triviality of the Gribov copies in the SDL gauges \cite{Dudal:2019bjh}, see Section \ref{ConformalSDL}. As a consequence of this triviality, using the Ward identities of the model---together with the symmetry between the topological and Faddeed-Popov ghosts introduced in \cite{Junqueira:2017zea}---and employing algebraic renormalization techniques, we conclude that $Z_g = 1$, \textit{i.e.}, that the self-dual BS $\beta$ function indeed vanishes. 

We observed that these theories do not possess the same quantum structure, by comparing the $\beta$ function of each model, see Section \ref{Beta}. From this analysis, we characterized the correspondence between the twisted $N=2$ SYM and BS theories at quantum level, defining in which regimes such a correspondence occurs, see Section \ref{DW/BS}. In spite of having distinct BRST constructions, we conclude that working in the instanton moduli space $\mathcal{A}/\mathcal{G}$ on the DW side is completely equivalent to working in the self-dual Landau gauges on the BS one, since the Gribov copies do not affect its infrared dynamics. In a few words, the twisted $N=2$ SYM in any Riemannian manifold (that can be continuously deformed into $M = \mathbb{R}^4$), in the weak coupling limit $g^2 \rightarrow 0$, is correspondent to the BS theory in the self-dual Landau gauges in an arbitrary perturbative regime, which consists of a conformal gauge theory defined in Euclidean flat space, see Table \ref{tableDW/BS}. Such a characterization could shed some light on the connection between supersymmetry, topology, off-shell BRST symmetry, and non-Abelian conformal gauge theories in four dimensions.  

\section*{Acknowledgments}

We would like to thank A. D. Pereira, G. Sadovski, and A. A. Tomaz for enlightening discussions, which was indispensable for the development of this work. This study was financed in part by The Coordena\c c\~ao de Aperfei\c coamento de Pessoal de N\'ivel Superior - Brasil (CAPES) -- Finance Code 001 -- and the Conselho Nacional de Desenvolvimento Científico e Tecnológico (CNPq-Brazil) -- Finance Code 159928/2019-2. 

\appendix

\section{BS Ward identities in the self-dual Landau gauges}\label{Ap1}

The BS action in the self-dual Landau gauges \eqref{fullaction} enjoys the following Ward identities:

(i) The Slavnov-Taylor identity, which expresses the BRST invariance of the action \eqref{fullaction}:
\begin{equation}
\mathcal{S}(\Sigma)=0\;,\label{st1}
\end{equation}
where
\begin{eqnarray}
\mathcal{S}(\Sigma)&=&\int d^4z\left[\left(\psi^a_\mu-\frac{\delta\Sigma}{\delta\Omega^a_\mu}\right)\frac{\delta\Sigma}{\delta A^a_\mu}+\frac{\delta\Sigma}{\delta\tau^a_\mu}\frac{\delta\Sigma}{\delta\psi^a_\mu}+\left(\phi^a+\frac{\delta\Sigma}{\delta L^a}\right)\frac{\delta\Sigma}{\delta c^a}+\frac{\delta\Sigma}{\delta E^a}\frac{\delta\Sigma}{\delta\phi^a}+\right.\nonumber\\
&+&\left.b^a\frac{\delta\Sigma}{\delta\bar{c}^a}+\bar{\eta}^a\frac{\delta\Sigma}{\delta\bar{\phi}^a}+B^a_{\mu\nu}\frac{\delta\Sigma}{\delta\bar{\chi}^a_{\mu\nu}}+\Omega^a_\mu\frac{\delta\Sigma}{\delta\tau^a_\mu}+L^a\frac{\delta\Sigma}{\delta E^a}+K^a_{\mu\nu}\frac{\delta\Sigma}{\delta\Lambda^a_{\mu\nu}}\right]\;.\label{st2}
\end{eqnarray}

(ii) Ordinary Landau gauge fixing and Faddeev-Popov anti-ghost equation:
\begin{eqnarray}
\mathcal{W}^a_{1} \Sigma &=& \frac{\delta\Sigma}{\delta b^a} =\partial_\mu A_\mu^a\;,\nonumber\\
\mathcal{W}^a_{2} \Sigma &=& \frac{\delta\Sigma}{\delta\bar{c}^a}-\partial_\mu\frac{\delta\Sigma}{\delta\Omega^a_\mu}=-\partial_\mu\psi_\mu^a\;.\label{Lgf1}
\end{eqnarray}

(iii) Topological Landau gauge fixing and bosonic anti-ghost equation:
\begin{eqnarray}
\mathcal{W}^a_{3} \Sigma &=&\frac{\delta\Sigma}{\delta\bar{\eta}^a}=\partial_\mu\psi_\mu^a\;,\nonumber\\
\mathcal{W}^a_{4} \Sigma &=& \frac{\delta\Sigma}{\delta\bar{\phi}^a}-\partial_\mu\frac{\delta\Sigma}{\delta\tau^a_\mu}= 0\;.\label{Lgf2}
\end{eqnarray}

(iv) Bosonic ghost equation:
\begin{equation}
\mathcal{G}^a_\phi\Sigma=\Delta_\phi^a\;,\label{bg1}
\end{equation}
where
\begin{eqnarray}
\mathcal{G}^a_\phi&=&\int d^4z\left(\frac{\delta}{\delta\phi^a}-gf^{abc}\bar{\phi}^b\frac{\delta}{\delta b^c}\right)\;,\nonumber\\
\Delta_\phi^a&=&gf^{abc}\int d^4z\left(\tau_\mu^bA_\mu^c+E^bc^c+\Lambda_{\mu\nu}^b\bar{\chi}_{\mu\nu}^c\right)\;.\label{bg2}
\end{eqnarray}

(v) Ordinary Faddeev-Popov ghost equation:
\begin{equation}
\mathcal{G}^a_1\Sigma=\Delta^a\;,\label{Og1}
\end{equation}
where
\begin{eqnarray}
\mathcal{G}^a_1&=&\int d^4z\left[\frac{\delta}{\delta c^a}+gf^{abc}\left(\bar{c}^b\frac{\delta}{\delta b^c}+\bar{\phi}^b\frac{\delta}{\delta\bar{\eta}^c}+\bar{\chi}^b_{\mu\nu}\frac{\delta}{\delta B^c_{\mu\nu}}+\Lambda^b_{\mu\nu}\frac{\delta}{\delta K^c_{\mu\nu}}\right)\right]\;,\nonumber\\
\Delta^a&=&gf^{abc}\int d^4z\left(E^b\phi^c-\Omega_\mu^bA_\mu^c-\tau_\mu^b\psi_\mu^c-L^bc^c+\Lambda_{\mu\nu}^bB_{\mu\nu}^c-K_{\mu\nu}^b\bar{\chi}_{\mu\nu}^c\right)\;.\label{Og2}
\end{eqnarray}

(vi) Second Faddeev-Popov ghost equation:
\begin{equation}
\mathcal{G}^a_2\Sigma=\Delta^a\;,\label{Sg1}
\end{equation}
where
\begin{equation}
\mathcal{G}^a_2=\int d^4z\left[\frac{\delta}{\delta c^a}-gf^{abc}\left(\bar{\phi}^b\frac{\delta}{\delta\bar{c}^c}+A^b_\mu\frac{\delta}{\delta\psi^c_\mu}+c^b\frac{\delta}{\delta\phi^c}-\bar{\eta}^b\frac{\delta}{\delta b^c}+E^b\frac{\delta}{\delta L^c}\right)\right]\;.\label{Sg2}
\end{equation}

(vii) Vector supersymmetry:
\begin{equation}
\mathcal{W}_\mu\Sigma=0\;,\label{w1}
\end{equation}
where
\begin{eqnarray}
\mathcal{W}_\mu &=&\int d^4z\left[\partial_\mu A_\nu^a\frac{\delta}{\delta \psi_\nu^a}+\partial_\mu c^a\frac{\delta}{\delta \phi^a}+\partial_\mu\bar{\chi}_{\nu\alpha}^a\frac{\delta}{\delta B_{\nu\alpha}^a}+\partial_\mu\bar{\phi}^a\left(\frac{\delta}{\delta\bar{\eta}^a}+\frac{\delta}{\delta\bar{c}^a}\right)+\right.\nonumber\\
&+&\left.\left(\partial_\mu\bar{c}^a-\partial_\mu\bar{\eta}^a\right)\frac{\delta}{\delta b^a}+\partial_\mu\tau_\nu^a\frac{\delta}{\delta \Omega_\nu^a}+\partial_\mu E^a\frac{\delta}{\delta L^a}+\partial_\mu\Lambda^a_{\nu\alpha}\frac{\delta}{\delta K_{\nu\alpha}^a}\right]\;.\label{w2}
\end{eqnarray}

(viii) Bosonic non-linear symmetry:
\begin{equation}
\mathcal{T}(\Sigma)=0\;,\label{cnl1}
\end{equation}
where
\begin{equation}
\mathcal{T}(\Sigma)=\int d^4z\left[\frac{\delta\Sigma}{\delta\Omega^a_\mu}\frac{\delta\Sigma}{\delta\psi^a_\mu}-\frac{\delta\Sigma}{\delta L^a}\frac{\delta\Sigma}{\delta\phi^a}-\frac{\delta\Sigma}{\delta K^a_{\mu\nu}}\frac{\delta\Sigma}{\delta B^a_{\mu\nu}}+\left(\bar{c}^a-\bar{\eta}^a\right)\left(\frac{\delta\Sigma}{\delta\bar{c}^a}+\frac{\delta\Sigma}{\delta\bar{\eta}^a}\right)\right]\;.\nonumber\\
\label{cnl2}
\end{equation}

(ix) Global ghost supersymmetry:
\begin{equation}
\mathcal{G}_3\Sigma=0\;,\label{gss1}
\end{equation}
where
\begin{equation}
\mathcal{G}_3=\int d^4z\left[\bar{\phi}^a\left(\frac{\delta}{\delta\bar{\eta}^a}+\frac{\delta}{\delta\bar{c}^a}\right)-c^a\frac{\delta}{\delta\phi^a}+\tau^a_\mu\frac{\delta}{\delta\Omega^a_\mu}+2E^a\frac{\delta}{\delta L^a}+\Lambda^a_{\mu\nu}\frac{\delta}{\delta K^a_{\mu\nu}}\right]\;.\label{gss2}
\end{equation}

The last two symmetries are the new ones introduced in \cite{Junqueira:2017zea}. The non-linear bosonic symmetry (vii) is precisely the one discussed in Section 3.2, see eq. \eqref{FPtopoghosts} which relates the FP and topological ghosts.  We remark that the Faddeev-Popov ghost equations \eqref{Og1} and \eqref{Sg1} can be combined to obtain an exact global supersymmetry,
\begin{equation}
\Delta\mathcal{G}^a\Sigma=0\;,\label{cg1}
\end{equation}
where
\begin{eqnarray}
\Delta\mathcal{G}^a&=&\mathcal{G}_1^a-\mathcal{G}_2^a\;=\;\int d^4z\;f^{abc}\left[\left(\bar{c}^b-\bar{\eta}^b\right)\frac{\delta}{\delta b^c}+\bar{\phi}^b\left(\frac{\delta}{\delta\bar{\eta}^c}+\frac{\delta}{\delta\bar{c}^c}\right)+A_\mu^b\frac{\delta}{\delta\psi_\mu^c}+\right.\nonumber\\
&+&\left.\bar{\chi}^b_{\mu\nu}\frac{\delta}{\delta B_{\mu\nu}^c}+c^b\frac{\delta}{\delta\phi^c}+\Lambda^b_{\mu\nu}\frac{\delta}{\delta K_{\mu\nu}^c}+\tau_\mu^b\frac{\delta}{\delta\Omega_\mu^c}+E^b\frac{\delta}{\delta L^c}\right]\;.\label{cg2}
\end{eqnarray}
We observe the similarity of the equation \eqref{cg1} with the vector supersymmetry \eqref{w1}. It is also worth mentioning that, even though the ghost number of the operator \eqref{cg2} is $-1$, resembling an anti-BRST symmetry, it is not a genuine anti-BRST symmetry---see for instance \cite{Braga:1999ui} for the explicit anti-BRST symmetry in topological gauge theories.

\bibliographystyle{utphys2}
\bibliography{library}

\providecommand{\href}[2]{#2}\begingroup\raggedright\begin{thebibliography}{10}

\bibitem{Belavin:1975fg}
A.~A. Belavin, A.~M. Polyakov, A.~S. Schwartz, and Y.~S. Tyupkin,
  ``{Pseudoparticle solutions of the Yang-Mills equations}''.
  \href{http://dx.doi.org/10.1016/0370-2693(75)90163-X}{{\em Physics Letters B}
  {\bfseries 59} no.~1, (1975) 85--87}.

\bibitem{DONALDSON1990257}
S.~Donaldson, ``Polynomial invariants for smooth four-manifolds''.
  \href{http://dx.doi.org/https://doi.org/10.1016/0040-9383(90)90001-Z}{{\em
  Topology} {\bfseries 29} no.~3, (1990) 257 -- 315}.
  \url{http://www.sciencedirect.com/science/article/pii/004093839090001Z}.

\bibitem{Donaldson:1983wm}
S.~K. Donaldson, ``{An application of gauge theory to four-dimensional
  topology}''. \href{http://dx.doi.org/10.4310/jdg/1214437665}{{\em Journal of
  Differential Geometry} {\bfseries 18} no.~2, (1983) 279--315}.

\bibitem{donaldson1987}
S.~K. Donaldson, ``The orientation of yang-mills moduli spaces and 4-manifold
  topology''. \href{http://dx.doi.org/10.4310/jdg/1214441485}{{\em J.
  Differential Geom.} {\bfseries 26} no.~3, (1987) 397--428}.
  \url{https://doi.org/10.4310/jdg/1214441485}.

\bibitem{floer1987}
A.~Floer, ``Morse theory for fixed points of symplectic diffeomorphisms''. {\em
  Bull. Amer. Math. Soc. (N.S.)} {\bfseries 16} no.~2, (04, 1987) 279--281.
  \url{https://projecteuclid.org:443/euclid.bams/1183553837}.

\bibitem{floer1988}
A.~Floer, ``An instanton-invariant for $3$-manifolds''. {\em Comm. Math. Phys.}
  {\bfseries 118} no.~2, (1988) 215--240.
  \url{https://projecteuclid.org:443/euclid.cmp/1104161987}.

\bibitem{Atiyah:1987ri}
M.~Atiyah, ``{NEW INVARIANTS OF THREE-DIMENSIONAL AND FOUR-DIMENSIONAL
  MANIFOLDS}''. {\em Proc. Symp. Pure Math.} {\bfseries 48} (1988) 285--299.

\bibitem{Witten:1988ze}
E.~Witten, ``{Topological quantum field theory}''.
  \href{http://dx.doi.org/10.1007/BF01223371}{{\em Communications in
  Mathematical Physics} {\bfseries 117} no.~3, (9, 1988) 353--386}.

\bibitem{West:1990}
P.~West, \href{http://dx.doi.org/10.1142/1002}{{\em {Introduction to
  Supersymmetry and Supergravity}}}.
\newblock World Scientific, 5, 1990.

\bibitem{Schwarz:1978cn}
A.~S. Schwarz, ``{The Partition Function of Degenerate Quadratic Functional and
  Ray-Singer Invariants}''. \href{http://dx.doi.org/10.1007/BF00406412}{{\em
  Lett. Math. Phys.} {\bfseries 2} (1978) 247--252}.

\bibitem{Ray:1973sb}
D.~Ray and I.~Singer, ``{Analytic torsion for complex manifolds}''.
  \href{http://dx.doi.org/10.2307/1970909}{{\em Annals Math.} {\bfseries 98}
  (1973) 154--177}.

\bibitem{Witten:1988hf}
E.~Witten, ``{Quantum field theory and the Jones polynomial}''.
  \href{http://dx.doi.org/10.1007/BF01217730}{{\em Communications in
  Mathematical Physics} {\bfseries 121} no.~3, (9, 1989) 351--399}.

\bibitem{Baulieu:1988xs}
L.~Baulieu and I.~Singer, ``{Topological Yang-Mills symmetry}''.
  \href{http://dx.doi.org/10.1016/0920-5632(88)90366-0}{{\em Nuclear Physics B
  - Proceedings Supplements} {\bfseries 5} no.~2, (12, 1988) 12--19}.

\bibitem{Becchi:1975nq}
C.~Becchi, A.~Rouet, and R.~Stora, ``{Renormalization of gauge theories}''.
  \href{http://dx.doi.org/10.1016/0003-4916(76)90156-1}{{\em Annals of Physics}
  {\bfseries 98} no.~2, (6, 1976) 287--321}.

\bibitem{Tyutin:1975qk}
I.~V. Tyutin, ``{Gauge Invariance in Field Theory and Statistical Physics in
  Operator Formalism}''.

\bibitem{Piguet:1995er}
O.~Piguet and S.~P. Sorella,
  \href{http://dx.doi.org/10.1007/978-3-540-49192-7}{{\em {Algebraic
  Renormalization}}}, vol.~28 of {\em Lecture Notes in Physics Monographs}.
\newblock Springer Berlin Heidelberg, Berlin, Heidelberg, 1995.

\bibitem{vanBaal:1989aw}
P.~Van~Baal, ``{An introduction to Topological Yang-Mills Theory}''. {\em Acta
  Physica Polonica} {\bfseries B21} no.~2, (1990) 73.

\bibitem{Witten:1998wy}
E.~Witten, ``{AdS / CFT correspondence and topological field theory}''.
  \href{http://dx.doi.org/10.1088/1126-6708/1998/12/012}{{\em JHEP} {\bfseries
  12} (1998) 012}.

\bibitem{BenettiGenolini:2017zmu}
P.~Benetti~Genolini, P.~Richmond, and J.~Sparks, ``{Topological AdS/CFT}''.
  \href{http://dx.doi.org/10.1007/JHEP12(2017)039}{{\em JHEP} {\bfseries 12}
  (2017) 039}.

\bibitem{Agrawal:2020xek}
P.~Agrawal, S.~Gukov, G.~Obied, and C.~Vafa, ``{Topological Gravity as the
  Early Phase of Our Universe}''.

\bibitem{Weis:1997kj}
M.~Weis, ``{Topological Aspects of Quantum Gravity}''.
  \href{http://arxiv.org/abs/hep-th/9806179}{{\ttfamily hep-th/9806179}}.

\bibitem{Delduc:1996yh}
F.~Delduc, N.~Maggiore, O.~Piguet, and S.~Wolf, ``{Note on constrained
  cohomology}''. \href{http://dx.doi.org/10.1016/0370-2693(96)00879-9}{{\em
  Phys. Lett. B} {\bfseries 385} (1996) 132--138}.

\bibitem{Boldo:2003jq}
I.~S. Boldo, C.~P. Constantinidis, O.~Piguet, M.~Lefranc, J.~L. Boldo, C.~P.
  Constantinidis, F.~Gieres, M.~Lefrancois, and O.~Piguet, ``{Observables in
  Topological Yang-Mills Theories}''.
  \href{http://dx.doi.org/10.1142/S0217751X0401777X}{{\em International Journal
  of Modern Physics A} {\bfseries 19} no.~17n18, (3, 2003) 2971--3004},
  \href{http://arxiv.org/abs/hep-th/0303053}{{\ttfamily hep-th/0303053}}.

\bibitem{Junqueira:2017zea}
O.~C. Junqueira, A.~D. Pereira, G.~Sadovski, R.~F. Sobreiro, and A.~A. Tomaz,
  ``{Topological Yang-Mills theories in self-dual and anti-self-dual Landau
  gauges revisited}''. \href{http://dx.doi.org/10.1103/PhysRevD.96.085008}{{\em
  Physical Review D} {\bfseries 96} no.~8, (10, 2017) 085008},
  \href{http://arxiv.org/abs/1707.06666}{{\ttfamily 1707.06666}}.

\bibitem{Junqueira:2018zxr}
O.~C. Junqueira, A.~D. Pereira, G.~Sadovski, R.~F. Sobreiro, and A.~A. Tomaz,
  ``{More about the renormalization properties of topological Yang-Mills
  theories}''. \href{http://dx.doi.org/10.1103/PhysRevD.98.105017}{{\em
  Physical Review D} {\bfseries 98} no.~10, (11, 2018) 105017},
  \href{http://arxiv.org/abs/1807.01517}{{\ttfamily 1807.01517}}.

\bibitem{Junqueira:2018xgl}
O.~C. Junqueira, A.~D. Pereira, G.~Sadovski, R.~F. Sobreiro, and A.~A. Tomaz,
  ``{Absence of radiative corrections in four-dimensional topological
  Yang-Mills theories}''.
  \href{http://dx.doi.org/10.1103/PhysRevD.98.021701}{{\em Physical Review D}
  {\bfseries 98} no.~2, (7, 2018) 21701},
  \href{http://arxiv.org/abs/1805.01850}{{\ttfamily 1805.01850}}.

\bibitem{Brandhuber:1994uf}
A.~Brandhuber, O.~Moritsch, M.~de~Oliveira, O.~Piguet, and M.~Schweda, ``{A
  renormalized supersymmetry in the topological Yang-Mills field theory}''.
  \href{http://dx.doi.org/10.1016/0550-3213(94)90102-3}{{\em Nuclear Physics B}
  {\bfseries 431} no.~1-2, (12, 1994) 173--190},
  \href{http://arxiv.org/abs/hep-th/9407105}{{\ttfamily hep-th/9407105}}.

\bibitem{Dudal:2019bjh}
D.~Dudal, C.~Felix, O.~Junqueira, D.~Montes, A.~Pereira, G.~Sadovski,
  R.~Sobreiro, and A.~Tomaz, ``{Infinitesimal Gribov copies in gauge-fixed
  topological Yang-Mills theories}''.
  \href{http://dx.doi.org/10.1016/j.physletb.2020.135531}{{\em Phys. Lett. B}
  {\bfseries 807} (2020) 135531}.

\bibitem{Sorella:1989ri}
S.~Sorella, ``{Algebraic Characterization of the Topological $\sigma$ Model}''.
  \href{http://dx.doi.org/10.1016/0370-2693(89)90527-3}{{\em Phys. Lett. B}
  {\bfseries 228} (1989) 69--74}.

\bibitem{Blasi:1989ka}
A.~Blasi and R.~Collina, ``{Basic Cohomology of Topological Quantum Field
  Theories}''. \href{http://dx.doi.org/10.1016/0370-2693(89)90336-5}{{\em Phys.
  Lett. B} {\bfseries 222} (1989) 419--424}.

\bibitem{Labastida:1997pb}
J.~M.~F. Labastida and C.~Lozano, ``{Lectures in Topological Quantum Field
  Theory}''. \href{http://arxiv.org/abs/hep-th/9709192}{{\ttfamily
  hep-th/9709192}}.

\bibitem{Kugo:1979gm}
T.~Kugo and I.~Ojima, ``{Local Covariant Operator Formalism of Nonabelian Gauge
  Theories and Quark Confinement Problem}''.
  \href{http://dx.doi.org/10.1143/PTPS.66.1}{{\em Prog. Theor. Phys. Suppl.}
  {\bfseries 66} (1979) 1--130}.

\bibitem{Wess:1992cp}
J.~Wess and J.~Bagger, {\em {Supersymmetry and supergravity}}.
\newblock Princeton University Press, Princeton, NJ, USA, 1992.

\bibitem{Blasi:2000qw}
A.~Blasi, V.~Lemes, N.~Maggiore, S.~Sorella, A.~Tanzini, O.~Ventura, and
  L.~Vilar, ``{Perturbative beta function of N=2 superYang-Mills theories}''.
  \href{http://dx.doi.org/10.1088/1126-6708/2000/05/039}{{\em JHEP} {\bfseries
  05} (2000) 039}.

\bibitem{Witten:1982im}
E.~Witten, ``{Supersymmetry and Morse theory}''. {\em J. Diff. Geom.}
  {\bfseries 17} no.~4, (1982) 661--692.

\bibitem{Maggiore:1994dw}
N.~Maggiore, ``{Algebraic renormalization of N=2 superYang-Mills theories
  coupled to matter}''. \href{http://dx.doi.org/10.1142/S0217751X95001789}{{\em
  Int. J. Mod. Phys. A} {\bfseries 10} (1995) 3781--3802}.

\bibitem{Atiyah:1963zz}
M.~F. Atiyah and I.~M. Singer, ``{The index of elliptic operators on compact
  manifolds}''.
\href{http://dx.doi.org/10.1090/S0002-9904-1963-10957-X}{{\em Bull. Am. Math.
  Soc.} {\bfseries 69} (1969) 422--433}.

\bibitem{Atiyah:1978wi}
M.~F. Atiyah, N.~J. Hitchin, and I.~M. Singer, ``{Self-Duality in
  Four-Dimensional Riemannian Geometry}''.
  \href{http://dx.doi.org/10.1098/rspa.1978.0143}{{\em Proceedings of the Royal
  Society A: Mathematical, Physical and Engineering Sciences} {\bfseries 362}
  no.~1711, (9, 1978) 425--461}.

\bibitem{Tong:2005un}
D.~Tong, ``{TASI lectures on solitons: Instantons, monopoles, vortices and
  kinks}''. in {\em {Theoretical Advanced Study Institute in Elementary
  Particle Physics}: {Many Dimensions of String Theory}}.
\newblock 6, 2005.

\bibitem{tHooft:1976snw}
G.~'t~Hooft, ``{Computation of the quantum effects due to a four-dimensional
  pseudoparticle}''. \href{http://dx.doi.org/10.1103/PhysRevD.14.3432}{{\em
  Physical Review D} {\bfseries 14} no.~12, (12, 1976) 3432--3450}.

\bibitem{DAdda:1977sqj}
A.~D'Adda and P.~Di~Vecchia, ``{Supersymmetry and Instantons}''.
  \href{http://dx.doi.org/10.1016/0370-2693(78)90826-2}{{\em Phys. Lett. B}
  {\bfseries 73} (1978) 162}.

\bibitem{Blau:1990nv}
M.~Blau and G.~Thompson, ``{Do metric independent classical actions lead to
  topological field theories?}''.
  \href{http://dx.doi.org/10.1016/0370-2693(91)90262-O}{{\em Physics Letters B}
  {\bfseries 255} no.~4, (2, 1991) 535--542}.

\bibitem{Abud:1991mu}
M.~Abud and G.~Fiore, ``{Batalin-Vilkovisky approach to the metric independence
  of TQFT}''. \href{http://dx.doi.org/10.1016/0370-2693(92)91484-Q}{{\em
  Physics Letters B} {\bfseries 293} no.~1-2, (10, 1992) 89--93}.

\bibitem{Mardones:1990qc}
A.~Mardones and J.~Zanelli, ``{Lovelock-Cartan theory of gravity}''.
  \href{http://dx.doi.org/10.1088/0264-9381/8/8/018}{{\em Classical and Quantum
  Gravity} {\bfseries 8} no.~8, (8, 1991) 1545--1558}.

\bibitem{Daniel:1979ez}
M.~Daniel and C.~M. Viallet, ``{The geometrical setting of gauge theories of
  the Yang-Mills type}''.
  \href{http://dx.doi.org/10.1103/RevModPhys.52.175}{{\em Reviews of Modern
  Physics} {\bfseries 52} no.~1, (1, 1980) 175--197}.

\bibitem{Vandersickel:2011zc}
N.~Vandersickel, {\em {A study of the Gribov-Zwanziger action: from propagators
  to glueballs}}.
\newblock PhD thesis, Ghent University, 4, 2011.
\newblock \href{http://arxiv.org/abs/1104.1315}{{\ttfamily 1104.1315}}.

\bibitem{Dixon:1979bs}
J.~A. Dixon,
``{COHOMOLOGY AND RENORMALIZATION OF GAUGE THEORIES. 2.}''.

\bibitem{Ouvry:1988mm}
S.~Ouvry, R.~Stora, and P.~Van~Baal, ``{On the algebraic characterization of
  Witten's topological Yang-Mills theory}''.
  \href{http://dx.doi.org/10.1016/0370-2693(89)90029-4}{{\em Physics Letters B}
  {\bfseries 220} no.~1-2, (1989) 159--163}.

\bibitem{Horne:1988yn}
J.~H. Horne, ``{Superspace Versions of Topological Theories}''.
  \href{http://dx.doi.org/10.1016/0550-3213(89)90046-1}{{\em Nucl. Phys. B}
  {\bfseries 318} (1989) 22--52}.

\bibitem{Kanno:1988wm}
H.~Kanno, ``{Weyl Algebra Structure and Geometrical Meaning of {BRST}
  Transformation in Topological Quantum Field Theory}''.
  \href{http://dx.doi.org/10.1007/BF01506544}{{\em Z. Phys. C} {\bfseries 43}
  (1989) 477}.

\bibitem{WerneckdeOliveira:1993pa}
M.~Werneck~de Oliveira, ``{Algebraic renormalization of the topological
  Yang-Mills field theory}''.
  \href{http://dx.doi.org/10.1016/0370-2693(93)90231-6}{{\em Phys. Lett. B}
  {\bfseries 307} (1993) 347--352}.

\bibitem{Brooks:1988jm}
R.~Brooks, D.~Montano, and J.~Sonnenschein, ``{Gauge fixing and renormalization
  in topological quantum field theory}''.
  \href{http://dx.doi.org/10.1016/0370-2693(88)90458-3}{{\em Physics Letters B}
  {\bfseries 214} no.~1, (11, 1988) 91--97}.

\bibitem{Lemes:2000ni}
V.~Lemes, N.~Maggiore, M.~Sarandy, S.~Sorella, A.~Tanzini, and O.~Ventura,
  ``{Nonrenormalization theorems for N=2 super-Yang-Mills}''.

\bibitem{Fucito:1997xm}
F.~Fucito, A.~Tanzini, L.~C.~Q. Vilar, O.~S. Ventura, C.~A.~G. Sasaki, and
  S.~P. Sorella, ``{Algebraic Renormalization: perturbative twisted
  considerations on topological Yang-Mills theory and on N=2 supersymmetric
  gauge theories}''. {\em 1st School on Field Theory and Gravitation Vitoria,
  Brazil, April 15-19, 1997} {\bfseries hep-th} (7, 1997) 15--19,
  \href{http://arxiv.org/abs/hep-th/9707209}{{\ttfamily hep-th/9707209}}.

\bibitem{Birmingham:1988ap}
D.~Birmingham, M.~Rakowski, G.~Thompson, I.~Centre, T.~Pto, T.~Physics,
  I.~Centre, T.~Pttvsk, P.~Vi, and P.~Jussieu, ``{Renormalization of
  topological field theory}''.
  \href{http://dx.doi.org/10.1016/0550-3213(90)90058-L}{{\em Nuclear Physics B}
  {\bfseries 329} no.~1, (1, 1990) 83--97}.

\bibitem{Batalin:1984jr}
I.~Batalin and G.~Vilkovisky, ``{Quantization of Gauge Theories with Linearly
  Dependent Generators}''.
  \href{http://dx.doi.org/10.1103/PhysRevD.28.2567}{{\em Phys. Rev. D}
  {\bfseries 28} (1983) 2567--2582}. [Erratum: Phys.Rev.D 30, 508 (1984)].

\bibitem{Witten:1976ck}
E.~Witten, ``{Some Exact Multi - Instanton Solutions of Classical Yang-Mills
  Theory}''. \href{http://dx.doi.org/10.1103/PhysRevLett.38.121}{{\em Phys.
  Rev. Lett.} {\bfseries 38} (1977) 121--124}.
[,124(1976)].

\bibitem{Jackiw:1976fs}
R.~Jackiw, C.~Nohl, and C.~Rebbi, ``{Conformal Properties of Pseudoparticle
  Configurations}''. \href{http://dx.doi.org/10.1103/PhysRevD.15.1642}{{\em
  Phys. Rev.} {\bfseries D15} (1977) 1642}.
[,128(1976)].

\bibitem{Gribov:1977wm}
V.~Gribov, ``{Quantization of non-Abelian gauge theories}''.
  \href{http://dx.doi.org/10.1016/0550-3213(78)90175-X}{{\em Nuclear Physics B}
  {\bfseries 139} no.~1-2, (6, 1978) 1--19}.

\bibitem{Sobreiro:2005ec}
R.~F. Sobreiro and S.~P. Sorella, ``{Introduction to the Gribov Ambiguities In
  Euclidean Yang-Mills Theories}''.
  \href{http://arxiv.org/abs/hep-th/0504095}{{\ttfamily hep-th/0504095}}.

\bibitem{Zwanziger:1988jt}
D.~Zwanziger, ``{Action From the Gribov Horizon}''.
  \href{http://dx.doi.org/10.1016/0550-3213(89)90263-0}{{\em Nucl. Phys. B}
  {\bfseries 321} (1989) 591--604}.

\bibitem{Singer:1978dk}
I.~M. Singer, ``{Some remarks on the Gribov ambiguity}''.
  \href{http://dx.doi.org/10.1007/BF01609471}{{\em Communications in
  Mathematical Physics} {\bfseries 60} no.~1, (2, 1978) 7--12}.

\bibitem{Faddeev:1967fc}
L.~Faddeev and V.~Popov, ``{Feynman diagrams for the Yang-Mills field}''.
  \href{http://dx.doi.org/10.1016/0370-2693(67)90067-6}{{\em Physics Letters B}
  {\bfseries 25} no.~1, (7, 1967) 29--30}.

\bibitem{Itzykson:1980rh}
C.~Itzykson and J.~Zuber, {\em {Quantum Field Theory}}.
\newblock International Series In Pure and Applied Physics. McGraw-Hill, New
  York, 1980.

\bibitem{Sobreiro:2005vn}
R.~Sobreiro and S.~Sorella, ``{A Study of the Gribov copies in linear covariant
  gauges in Euclidean Yang-Mills theories}''.
  \href{http://dx.doi.org/10.1088/1126-6708/2005/06/054}{{\em JHEP} {\bfseries
  06} (2005) 054}.

\bibitem{Capri:2015pja}
M.~Capri, A.~Pereira, R.~Sobreiro, and S.~Sorella, ``{Non-perturbative
  treatment of the linear covariant gauges by taking into account the Gribov
  copies}''. \href{http://dx.doi.org/10.1140/epjc/s10052-015-3707-z}{{\em Eur.
  Phys. J. C} {\bfseries 75} no.~10, (2015) 479}.

\bibitem{Capri:2015ixa}
M.~Capri, D.~Dudal, D.~Fiorentini, M.~Guimaraes, I.~Justo, A.~Pereira,
  B.~Mintz, L.~Palhares, R.~Sobreiro, and S.~Sorella, ``{Exact nilpotent
  nonperturbative BRST symmetry for the Gribov-Zwanziger action in the linear
  covariant gauge}''. \href{http://dx.doi.org/10.1103/PhysRevD.92.045039}{{\em
  Phys. Rev. D} {\bfseries 92} no.~4, (2015) 045039}.

\bibitem{Capri:2015nzw}
M.~Capri, D.~Fiorentini, M.~Guimaraes, B.~Mintz, L.~Palhares, S.~Sorella,
  D.~Dudal, I.~Justo, A.~Pereira, and R.~Sobreiro, ``{More on the
  nonperturbative Gribov-Zwanziger quantization of linear covariant gauges}''.
  \href{http://dx.doi.org/10.1103/PhysRevD.93.065019}{{\em Phys. Rev. D}
  {\bfseries 93} no.~6, (2016) 065019}.

\bibitem{DellAntonio:1991mms}
G.~Dell'Antonio and D.~Zwanziger, ``{Every gauge orbit passes inside the Gribov
  horizon}''. \href{http://dx.doi.org/10.1007/BF02099494}{{\em Communications
  in Mathematical Physics} {\bfseries 138} no.~2, (5, 1991) 291--299}.

\bibitem{Vandersickel:2010ti}
N.~Vandersickel, D.~Dudal, O.~Oliveira, and S.~P. Sorella, ``{From propagators
  to glueballs in the Gribov-Zwanziger framework}''.
  \href{http://dx.doi.org/10.1063/1.3574961}{{\em AIP Conf. Proc.} {\bfseries
  1343} (2011) 155--157}.

\bibitem{Gomez:2009tj}
A.~J. Gomez, M.~S. Guimaraes, R.~F. Sobreiro, and S.~P. Sorella, ``{Equivalence
  between Zwanziger's horizon function and Gribov's no-pole ghost form
  factor}''. \href{http://dx.doi.org/10.1016/j.physletb.2009.12.001}{{\em
  Physics Letters, Section B: Nuclear, Elementary Particle and High-Energy
  Physics} {\bfseries 683} no.~2-3, (10, 2009) 217--221},
  \href{http://arxiv.org/abs/0910.3596}{{\ttfamily 0910.3596}}.

\bibitem{Capri:2012wx}
M.~Capri, D.~Dudal, M.~Guimaraes, L.~Palhares, and S.~Sorella, ``{An all-order
  proof of the equivalence between Gribov's no-pole and Zwanziger's horizon
  conditions}''. \href{http://dx.doi.org/10.1016/j.physletb.2013.01.039}{{\em
  Physics Letters B} {\bfseries 719} no.~4-5, (2, 2013) 448--453},
  \href{http://arxiv.org/abs/1212.2419}{{\ttfamily 1212.2419}}.

\bibitem{Dudal:2008sp}
D.~Dudal, J.~A. Gracey, S.~P. Sorella, N.~Vandersickel, and H.~Verschelde, ``{A
  Refinement of the Gribov-Zwanziger approach in the Landau gauge: Infrared
  propagators in harmony with the lattice results}''.
  \href{http://dx.doi.org/10.1103/PhysRevD.78.065047}{{\em Phys. Rev. D}
  {\bfseries 78} (2008) 065047}.

\bibitem{Slavnov:1989jh}
A.~A. Slavnov, ``{Physical Unitarity in the \{BRST\} Approach}''.
  \href{http://dx.doi.org/10.1016/0370-2693(89)91521-9}{{\em Phys. Lett. B}
  {\bfseries 217} (1989) 91--94}.

\bibitem{Frolov:1989az}
S.~A. Frolov and A.~A. Slavnov, ``{Construction of the Effective Action for
  General Gauge Theories via Unitarity}''.
  \href{http://dx.doi.org/10.1016/0550-3213(90)90562-R}{{\em Nucl. Phys. B}
  {\bfseries 347} (1990) 333--346}.

\bibitem{Dudal:2009xh}
D.~Dudal, S.~P. Sorella, N.~Vandersickel, and H.~Verschelde, ``{Gribov no-pole
  condition, Zwanziger horizon function, Kugo-Ojima confinement criterion,
  boundary conditions, BRST breaking and all that}''.
  \href{http://dx.doi.org/10.1103/PhysRevD.79.121701}{{\em Phys. Rev. D}
  {\bfseries 79} (2009) 121701}.

\bibitem{Baulieu:2008fy}
L.~Baulieu and S.~P. Sorella, ``{Soft breaking of BRST invariance for
  introducing non-perturbative infrared effects in a local and renormalizable
  way}''. \href{http://dx.doi.org/10.1016/j.physletb.2008.11.036}{{\em Phys.
  Lett.} {\bfseries B671} no.~4-5, (8, 2008) 481--485},
  \href{http://arxiv.org/abs/0808.1356}{{\ttfamily 0808.1356}}.

\bibitem{Sorella:2009vt}
S.~P. Sorella, ``{Gribov horizon and BRST symmetry: A Few remarks}''.
  \href{http://dx.doi.org/10.1103/PhysRevD.80.025013}{{\em Phys. Rev. D}
  {\bfseries 80} (2009) 025013}.

\bibitem{Sorella:2011tu}
S.~P. Sorella, D.~Dudal, M.~S. Guimaraes, and N.~Vandersickel, ``{Features of
  the Refined Gribov-Zwanziger theory: Propagators, BRST soft symmetry breaking
  and glueball masses}''. \href{http://dx.doi.org/10.22323/1.117.0022}{{\em
  PoS} {\bfseries FACESQCD} (2010) 022}.

\bibitem{Capri:2016aqq}
M.~A.~L. Capri, D.~Dudal, D.~Fiorentini, M.~S. Guimaraes, I.~F. Justo, A.~D.
  Pereira, B.~W. Mintz, L.~F. Palhares, R.~F. Sobreiro, and S.~P. Sorella,
  ``{Local and BRST-invariant Yang-Mills theory within the Gribov horizon}''.
\href{http://dx.doi.org/10.1103/PhysRevD.94.025035}{{\em Phys. Rev.} {\bfseries
  D94} no.~2, (2016) 025035}.

\bibitem{Pereira:2016fpn}
A.~D. Pereira, R.~F. Sobreiro, and S.~P. Sorella, ``{Non-perturbative BRST
  quantization of Euclidean Yang\textendash{}Mills theories in
  Curci\textendash{}Ferrari gauges}''.
  \href{http://dx.doi.org/10.1140/epjc/s10052-016-4368-2}{{\em Eur. Phys. J. C}
  {\bfseries 76} no.~10, (2016) 528}.

\bibitem{Capri:2018ijg}
M.~A.~L. Capri, D.~Dudal, M.~S. Guimaraes, A.~D. Pereira, B.~W. Mintz, L.~F.
  Palhares, and S.~P. Sorella, ``{The universal character of Zwanziger's
  horizon function in Euclidean Yang–Mills theories}''.
\href{http://dx.doi.org/10.1016/j.physletb.2018.03.058}{{\em Phys. Lett.}
  {\bfseries B781} (2018) 48--54}.

\bibitem{Capri:2014tta}
M.~A.~L. Capri, M.~S. Guimaraes, I.~F. Justo, L.~F. Palhares, and S.~P.
  Sorella, ``{On the irrelevance of the Gribov issue in $\mathcal{N}=4$ Super
  Yang-Mills in the Landau gauge}''.
  \href{http://dx.doi.org/10.1016/j.physletb.2014.06.035}{{\em Physics Letters
  B} {\bfseries 735} (4, 2014) 277--281},
  \href{http://arxiv.org/abs/1404.7163}{{\ttfamily 1404.7163}}.

\bibitem{Boldo:2003ci}
J.~L. Boldo, C.~P. Constantinidis, O.~Piguet, F.~Gieres, and
  M.~Lefran{\c{c}}ois, ``{Topological Yang-Mills Theories and their
  Observables: A Superspace Approach}''.
  \href{http://dx.doi.org/10.1142/S0217751X03015568}{{\em International Journal
  of Modern Physics A} {\bfseries 18} no.~12, (5, 2003) 2119--2125},
  \href{http://arxiv.org/abs/hep-th/0303084}{{\ttfamily hep-th/0303084}}.

\bibitem{Braga:1999ui}
N.~R. Braga and C.~F. Godinho, ``{Extended BRST invariance in topological
  Yang-Mills theory revisited}''.
  \href{http://dx.doi.org/10.1103/PhysRevD.61.125019}{{\em Phys. Rev. D}
  {\bfseries 61} (2000) 125019}.

\end{thebibliography}\endgroup

\end{document}